\documentclass[twocolumn,tighten]{aastex631}

\usepackage{color,comment,multirow,booktabs,upgreek}
\usepackage[hang,flushmargin]{footmisc} 

\graphicspath{{./}{images/}}

\newcommand{\hi}{\ifmmode{\rm HI}\else{H\/{\sc i}}\fi} 
\newcommand{\vlos}{\ifmmode{V_\mathrm{los}}\else{$V_\mathrm{los}$}\fi}
\newcommand{\vsys}{\ifmmode{V_\mathrm{sys}}\else{$V_\mathrm{sys}$}\fi}
\newcommand{\vrot}{\ifmmode{V_\mathrm{rot}}\else{$V_\mathrm{rot}$}\fi}
\newcommand{\vrad}{\ifmmode{V_\mathrm{rad}}\else{$V_\mathrm{rad}$}\fi}
\newcommand{\vdisp}{\ifmmode{\sigma_\mathrm{gas}}\else{$\sigma_\mathrm{gas}$}\fi}
\newcommand{\vflat}{\ifmmode{V_\mathrm{flat}}\else{$V_\mathrm{flat}$}\fi} 
\newcommand{\de}{\ifmmode{^\circ}\else{$^\circ$}\fi} 
\newcommand {\kms}{\ifmmode{\rm km \, s^{-1}}\else{$\rm km \, s^{-1}$}\fi} 
\newcommand {\mo}{{\rm M}_\odot}

\newcommand {\moyr}{\,{\rm M_\odot\,\rm yr}^{-1}}
\newcommand {\mdot}{\dot{M}}
\newcommand {\mdotaver}{\langle \, \mdot \, \rangle}

\newcommand{\eqref}[1]{Eq.\ (\ref{#1})}

\newcommand{\mstar}{\ifmmode{M_{\star}}\else{$M_{\star}$}\fi}
\newcommand{\bba}{$^{\rm 3D}$B\textsc{arolo}}
\newcommand{\galnum}{54}

\shorttitle{Radial flows in local galaxies}
\shortauthors{Di Teodoro et al.}

\begin{document}

\title{\large Radial motions and radial gas flows in local spiral galaxies}%

\correspondingauthor{E.~M. Di Teodoro}
\email{editeodoro@jhu.edu}

\author[0000-0003-4019-0673]{Enrico M.\ Di Teodoro}
\affiliation{Department of Physics \& Astronomy, Johns Hopkins University, Baltimore, MD 21218, USA}
\affiliation{Space Telescope Science Institute, 3700 San Martin Drive, Baltimore, MD 21218, USA}

\author[0000-0003-4797-7030]{J.\ E.\ G.\ Peek}
\affiliation{Space Telescope Science Institute, 3700 San Martin Drive, Baltimore, MD 21218, USA}
\affiliation{Department of Physics \& Astronomy, Johns Hopkins University, Baltimore, MD 21218, USA}

\begin{abstract}
We determine radial velocities and mass flow rates in a sample of \galnum\ local spiral galaxies by modelling high-resolution and high-sensitivity data of the atomic hydrogen emission line. 
We found that, although radial inflow motions seem to be slightly preferred over outflow motions, their magnitude is generally small. 
Most galaxies show radial flows of only a few $\kms$ throughout their \hi\ disks, either inwards or outwards, without any clear increase in magnitude in the outermost regions, as we would expect for continuous radial accretion. Gas mass flow rates for most galaxies are less than $1 \; \moyr$.
Over the entire sample, we estimated an average inflow rate of $0.3\, \moyr$ outside the optical disk and of $0.1\, \moyr$ in the outskirts of the \hi\ disks. 
These inflow rates are about $5-10$ times smaller than the average star formation rate of $1.4\, \moyr$.
Our study suggests that there is no clear evidence for systematic radial accretion inflows that alone could feed and sustain the star formation process in the inner regions of local spiral galaxies at its current rate. 
\end{abstract}


\section{Introduction}
\label{sec:intro}
Gas accretion is an important process in the evolution of galaxies. 
Most star-forming spiral galaxies, including our own Milky Way, are believed to have been forming stars at an almost constant, or only slightly declining, rate over cosmic time \citep[e.g.,][]{Panter+2007,Aumer+2009,Madau+2014}. 
Given their average amount of gas, star-forming galaxies should deplete their entire gas content on timescales of only a few billion years \citep[e.g.][]{Leroy+2008,Bigiel+2011}.
To sustain their star-formation rate (SFR) at a quasi-steady state for longer times, a continuous replenishment of fresh gas from the surroundings at a rate roughly equal to the SFR is needed \citep[e.g.,][]{Fraternali+2012}.
The need for continuous accretion of metal-poor gas is also invoked to explain the relative scarcity of low metallicity stars in disks \citep[G-dwarf problem,][]{vandenBergh+1962,Worthey+1996,Haywood+2019}.

Fresh gas can be acquired either via galaxy merging or via accretion from the circumgalactic and intergalactic media (CGM and IGM, respectively).
Observational evidence suggests that, at the least in the local Universe, mergers can not bring enough gas into spiral galaxies \citep{DiTeodoro+2014}, while direct observation of gas accretion from the CGM/IGM still remains sparse and accretion rates very uncertain \citep{Sancisi+2008,Putman+2012}.
From a theoretical point of view, hydrodynamical simulations of galaxy formation
predict that large amounts of pristine gas can be accreted onto galaxies from the IGM, either via cold gas filaments that directly feed the inner regions of a galaxy (``cold'' mode) or via cooling of hot coronal gas that surrounds galactic disks (``hot'' mode) \citep[e.g.,][]{Keres+2005,Nelson+2013,Stern+2020}. 
According to simulations, gas accretion occurs mostly through the cold mode in low mass galaxies and at high redshifts, while the hot mode prevails in more massive systems and at low redshift.

In the classical picture of galaxy formation, disk galaxies grow inside-out over time by acquiring high angular momentum material from the large-structure IGM \citep[e.g.,][]{Fall+1980,Pichon+2011,Lagos+2017,El-Badry+2018}. 
Gas infalling onto a galaxy, subject to the torques exerted by the rotating disk/halo, can settle into a co-rotating accretion disk that extend out to very large radii \citep{Stewart+2011,Stewart+2013}. 
Therefore, cosmological accretion is expected to occur mostly in the outskirts of galactic disks. 
The presence of large co-rotating disks or torus-like structures of cool material has been also proposed to explain observations of the CGM in low-redshift star-forming galaxies \citep[e.g.,][]{Tumlinson+2017,Ho+2017}.
To become available for star formation, this fresh gas needs to be transported efficiently from the outermost to the inner regions of a disk, which is where most of the star formation process occurs. 
In this picture, inward radial gas flows through a galactic disk are expected \citep[e.g.,][]{Ho+2019}. 
These radial flows are also believed to play a fundamental role in the chemical evolution of galaxies, in particular in the development of abundance gradients across the disk \citep[][among many others]{Lacey+1985,Goetz+1992,Portinari+2000,Mott+2013}, although the radial velocities needed by chemo-dynamical models are generally small, of the order of a few $\kms$ only \citep[e.g.][]{Bilitewski+2012,Pezzulli+2016}.

The atomic hydrogen (\hi) emission line at 21 cm is arguably the best way to trace the outskirts of gaseous disks of local spiral galaxies and search for these alleged radial gas inflows. 
Atomic gas dominates at large radii and \hi\ disks typically extend 2-3 times further away than stellar disks. 
The gas surface density in the outskirts of disks is low enough that a significant radial accretion would imply detectable radial inward velocities.
Moreover, \hi\ gas often show significant asymmetries in their outermost regions that might be a signature of on-going or recent gas accretion \citep{Sancisi+2008} from either the IGM or from mergers. 
This intrinsic disturbed nature of \hi\ disks makes it challenging to measure robustly the amount of mass that flows inward/outward.
A further complication arises from the fact that radial motions can easily be an order of magnitude or more smaller than typical rotation motions, which always dominate the overall kinematics of a disk galaxy.
Therefore, extracting the radial flow signal from spectral \hi\ datacubes is not trivial.

To date, a few studies have used \hi\ data to investigate radial motions in small samples of galaxies \citep[][]{Wong+2004,Trachternach+2008,Schmidt+2016}.
It is however still unclear whether radial inflows in spiral galaxies are negligible \citep{Wong+2004} or if, at least for a good fraction of galaxies, radial mass inflows can be large enough to feed and sustain the star formation \citep{Schmidt+2016}. 
In this work, we intend to improve upon previous studies by performing a systematic search for radial flows in local disks. 
We put together a relatively large sample of \galnum\ galaxies with high-quality \hi\ archival data and, using a state-of-the-art 3D modelling technique, we accurately derive their kinematics and we quantify the velocity and amount of gas that flows radially through their disks.

The remainder of this paper is organized as follows. 
\autoref{sec:data} introduces our galaxy sample and the data that we use in our analysis.
In \autoref{sec:methods}, we illustrate the expected signature of radial motions in a rotating disk, we describe in detail our kinematical modelling procedure and we test its robustness. 
The results of our analysis are presented in \autoref{sec:results}. \autoref{sec:disc} compares our findings to previous works and discusses the limitations of our measurements and implications for gas accretion theories. 
Finally, we summarize and conclude in \autoref{sec:concl}.

\section{Galaxy sample and data}
\label{sec:data}
For this study, we made use of publicly-available \hi\ interferometric data of local spiral galaxies. 
In particular, we chose to analyse a sample of nearby galaxies observed by some of the best available \hi\ surveys, including The \hi\ nearby Galaxy Survey \citep[THINGS,][]{Walter+2008}, the Hydrogen Accretion in LOcal GAlaxieS survey \citep[HALOGAS,][]{Heald+2011}, the Local Volume \hi\ Survey \citep[LVHIS,][]{Koribalski+2018}, the Westerbork survey of neutral Hydrogen in Irregular and SPiral galaxies \citep[WHISP,][]{vanderHulst+2001}, the Very Large Array (VLA) Imaging of Virgo in Atomic gas survey \citep[VIVA,][]{Chung+2009}, the  ATLAS$^{3\mathrm{D}}$ \hi\ survey \citep{Serra+2012} and the \hi\ eXtreme galaxy survey \citep[HIX,][]{Lutz+2017}.
Our sample was chosen on the basis of four main selection criteria: i) rotation velocity, ii) sensitivity of the \hi\ data, iii) number of resolution elements and iv) the inclination angle.  
Because we aim to study radial accretion flows in star-forming disks, we select only galaxies that reach a velocity of at least 100 $\kms$ in the flat part of their rotation curve. 
Measuring radial gas flows out to the outermost regions of disks requires very high-quality data, having both high sensitivity and high spatial resolution.
For this reason, we only select data with a \hi\ column-density sensitivity better than $\sim5\times10^{19}$ cm$^{-2}$ and we require that galaxies are detected with at least 15 spatial resolution elements across the major axis of the disk. 
Finally, to avoid problems with the derivation of galaxy kinematics, we discard high-inclined and low-inclined galaxies and we only accept disks with intermediate inclination angles $i$, i.e.\ $30\de\lesssim i \lesssim 80\de$.

\begin{figure*}[t]
\center
\includegraphics[width=0.99\textwidth]{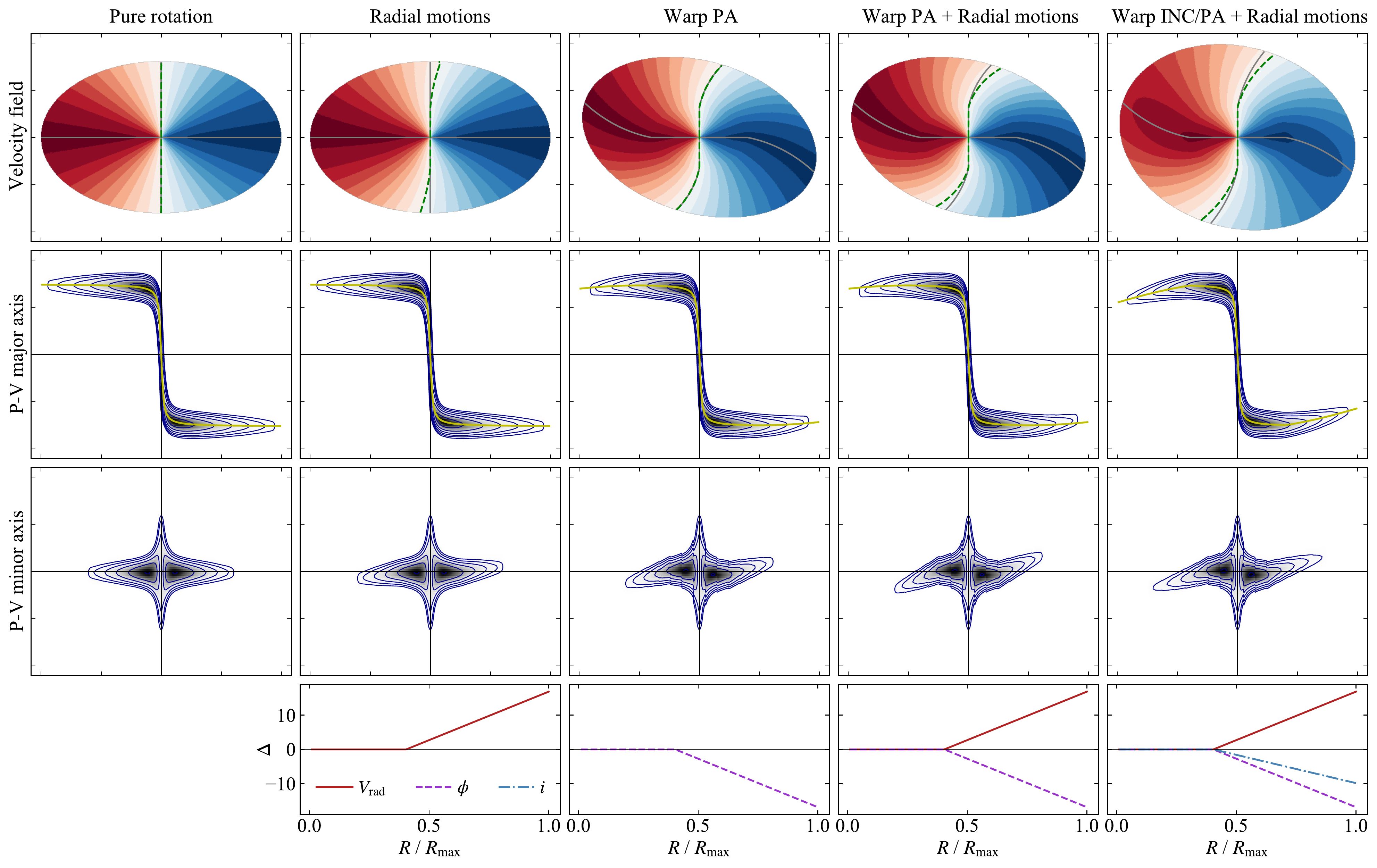}
\caption{Toy models of a rotating disk with warps and/or radial motions. From the left to the right: 1) a purely rotating disk with a flat rotation curve, 2) a disk with radial motions in the outer regions, 3) a disk with a warp in position angle in the outer regions, 4) a disk with both radial motions and a position angle warp and 5) a disk with radial motions and a warp in both position and inclination angle. We show the velocity field (first row), position-velocity slices along the major and minor axis (second and third rows, respectively) and the radial variations of \vrad\ ($\kms$, red solid line), $\phi$ (deg, purple dashed line) and $i$ (deg, cyan dashed-dotted line) with respect to their initial value in each model. On the velocity fields, the major/minor axes of the disk (grey solid line) and the velocity corresponding to the systemic velocity (green dashed line) are shown. On the P-V along the major axis, we plot the \vlos\ (yellow line), i.e.\ Eq.~\ref{eq:vlos} with $\theta=0\de$.}
\label{fig:toymodels}
\end{figure*} 

Our final sample is composed of \galnum\ well-known and well-studied spiral galaxies, that we list in \autoref{app:sample}. 
Our galaxies have distances $D\lesssim40$ kpc, stellar masses $9 \lesssim \log \mstar/\mo \lesssim 11$ and morphology ranging from early-type disks (e.g.\ S0, Sa) to late-type disks (e.g.\ Sb, Sc). 
Galaxies inhabit a range of different environments, from low-density environments to galaxy clusters (e.g. Virgo cluster).
Twelve galaxies are from THINGS, 8 from HALOGAS, 8 from WHISP, 6 from VIVA, 5 from ATLAS$^{3\mathrm{D}}$, 5 from LVHIS, 3 from HIX, while the remaining 7 are from different published studies (see \autoref{app:sample}).
Because we value sensitivity over spatial resolution for this study, we use natural-weighted instead of robust-weighted \citep{Briggs95} datacubes whenever available (e.g.\ THINGS, LVHIS). 
For WHISP data, we use datacubes smoothed to a 30$''$ spatial resolution.
\hi\ data for these galaxies have typical beam sizes $10''-100''$ and velocity channel widths of $4-8$ $\kms$.

A wealth of ancillary data and measurements are available for most of these galaxies in the literature.
For our purposes, we only need information on the distances and on the SFRs of our galaxies.
We adopt homogenized distances and uncertainties from the Extragalactic Distance Database \citep[EDD,][]{Tully+2009} and the CosmicFlows project \citep{Tully+2016}.
Star formation rates for 45 out 54 galaxies are taken from the catalog by \citet{Leroy+2019}, which were estimated combining data from the Wide-field Infrared Survey Explorer \citep[WISE,][]{Wright+2010} and from the Galaxy Evolution Explorer \citep[GALEX,][]{Martin+2005}.
For the remaining 9 galaxies, we estimate SFRs from WISE $W3$-band luminosities at 12 $\upmu$m, following the calibrations of \citet{Cluver+2017}. 
SFRs for our galaxies range from 0.01 $\moyr$ up to 6 $\moyr$, with average and median values over the entire sample of 1.4 and 0.9 $\moyr$, respectively. The typical uncertainty on the SFRs is 0.2 dex.
Finally, all galaxies in our sample have multi-band optical images from either the Sloan Digital Sky Survey \citep[SDSS,][]{Eisenstein+2011}, the Panoramic Survey Telescope and Rapid Response System survey \citep[Pan-STARRS,][]{Chambers+2016}, or the Dark Energy Spectroscopic Instrument (DESI) Legacy Surveys \citep[Legacy,][]{Dey+2019}.
We use these images to determine the central coordinates of our galaxies, the size of their stellar disks and whether the galaxies rotate clockwise or counter-clockwise.

\section{Methods}
\label{sec:methods}
\subsection{Radial motions in a rotating disk}
The first step in our analysis consists in deriving reliable radial velocities as a function of radius for our galaxies. 
The presence of radial motions in a galaxy modifies the emission-line profiles with respect to the case of a purely rotating disk. 
The geometry of a disk galaxy observed onto the plane of the sky is commonly parametrized through three quantities: 1) the center of the galaxy $(x_0,y_0)$, 2) the inclination angle $i$ of the midplane of the disk with respect to the line of sight ($i=90\de$ for edge-on) and 3) the position angle $\phi$ of the apparent major axis of the disk projected onto the sky.
A particle of gas subject to both rotation and radial motions in such a system will have an observed line-of-sight velocity \vlos\ given by \citep[e.g.,][]{Begeman87}:

\begin{equation}
    \label{eq:vlos}
    \vlos = \vsys + (\vrot\cos\theta + \vrad\sin\theta)\sin i
\end{equation}

\noindent where \vsys\ is the systemic velocity of the galaxy, \vrot\ is the azimuthal (rotation) velocity component, \vrad\ the radial velocity component and $\theta$ is the azimuthal angle into the plane of the disk, with $\theta=0\de$ corresponding to the major axis. 
Motions perpendicular to the plane of the disk are not considered here.
From Eq.~\ref{eq:vlos}, it is already clear that rotation has a larger effect on \vlos\ in the regions close to the major axis of a galaxy ($\theta\simeq0\de$), while radial motions show up more prominently on the minor axis ($\theta\simeq90\de$).

All above geometrical and kinematical parameters are usually not constant across a disk, but they can vary as a function of radius $R$. 
In particular, variations (warps) of the position and inclination angles, which are often observed in \hi\ disks \citep[e.g.,][]{Battaglia+2006}, can add complexity to the observed velocity structure of a galaxy.
We take advantage of \autoref{fig:toymodels} to illustrate the expected effect of radial motions and warps on an emission-line datacube of a galaxy.
Each column in \autoref{fig:toymodels} denotes a different kinematical toy model. 
We show a disk in pure rotation with a flat rotation curve as a reference, and the same disk with additional radial motions and/or a warp in the outer regions of the galaxy.
For each model, we plot the velocity field with kinematical axes (grey) and systemic velocity (green) highlighted, and position-velocity (P-V) slices taken through the galaxy center and along the horizontal and vertical direction, representing approximately (or exactly in cases with no warp) the major and minor axes of the disk, respectively. 
Both radial motions and a warp in the position angle (second and third model) alone cause a clear twist of the line-of-sight velocity contours near the minor axis with respect to a pure rotating disk.
However, while a warp produces an actual bending of the kinematical minor axis, resulting into an equivalent bend of the major axis as well, radial motions only introduce an offset between the systemic velocity and the observed velocity along the minor axis and have no effect on the major axis (Eq.~\ref{eq:vlos}).
As a consequence, when both a position angle warp and radial motions are present (fourth model), the former sets the twist of the major/minor axis, while the latter adds a further shift of the contour corresponding to the systemic velocity near the minor axis. 
A warp in inclination (last model) does not cause any further twist of the iso-velocity contours, but only scales the magnitude of the line-of-sight velocity and changes the projected shape of the disk.
The same effects are also clearly recognizable in the P-Vs along the major and minor axes. 
Because radial motions and warps have different effects in different regions of a disk, it is possible, although not trivial, to disentangle their relative contributions to the observed \vlos\ through kinematical model.

\subsection{3D tilted-ring modelling}
\label{sec:kinmod}

We used a custom-modified version of the 3D kinematical software \bba\footnote{Code is available at \href{https://editeodoro.github.io/Bbarolo}{https://editeodoro.github.io/Bbarolo}. This paper uses version 1.6.1. \citep{DiTeodoroBB} } \citep[BB, hereinafter][]{DiTeodoro&Fraternali15} to model the kinematics of our galaxy sample and to derive radial velocity profiles.
Unlike traditional methods that mostly apply to 2D velocity fields \citep[e.g.,][]{Begeman87, Spekkens+2007}, BB simulates \hi\ datacubes of galaxies using tilted-ring models and performs the fit directly in the 3D observational space (\textsc{3dfit} task). 
This approach exploits the full information available in 3D datasets, properly taking into account observational biases due to the finite spatial/spectral resolution of a telescope \citep[see][for details]{DiTeodoro&Fraternali15}. 
A galaxy model in BB is defined through four main geometrical parameters, $\mathbf{x}_\mathrm{geom} = (x_0,y_0,\phi,i)$, and four kinematical parameters, $\mathbf{x}_\mathrm{kin} = (\vsys,\vrot,\vrad,\vdisp)$, where \vdisp\ is the gas intrinsic velocity dispersion.
All these parameters can in theory be varied as a function of radius $R$.
Galaxy models defined in this way are mapped into a 3D emission-line datacube, having two spatial dimensions representing the position onto the plane of the sky $(x,y)$ and one spectral dimension representing the line-of-sight velocity \vlos, and the parameters optimized to reproduce the observed datacubes.

We used a standardized procedure and options to model the \galnum\ galaxies in our sample. 
First of all, to reduce the dimensionality of the problem, we decided to fix the center of galaxies and their systemic velocity during the fit. 
Because the spatial resolution of \hi\ data is much poorer than in optical data, we assumed galaxy centers taken from $z$-band optical images from the SDSS/Pan-STARRS/Legacy surveys. 
Uncertainties associated with optical centers are less than a few arcsecs, always much smaller than the beam size in the \hi\ data. 
Systemic velocities were calculated directly by BB from the global \hi\ profiles as the midpoint between the velocities corresponding to the 20\% of the peak flux, with typical errors of the order of the channel width. 
All others quantities $(\phi,i,\vrot,\vrad,\vdisp)$ are free parameters of our models and they are fitted to the data in a set of concentric rings at increasing radii.
The initial values for these parameters needed by the fitting algorithm are estimated by BB itself. 
Masks for the fit were built using the source-finder method, with a primary threshold of $5\times\sigma_\mathrm{rms}$ and a secondary threshold of $2\times\sigma_\mathrm{rms}$, where $\sigma_\mathrm{rms}$ is the root-mean-square noise of the data (see BB documentation and main paper for further details). 

After extensive testing on mock observations (see \autoref{sec:tests}), we determined that a three-step fit approach allowed us to recover robust radial velocity profiles, disentangling the effect of warps and radial motions. 
The first two steps aim to determine the rotation curve and the warp and follow a classical tilted-ring modelling procedure: during an initial fit, radial motions are null while the geometrical and the other kinematical parameters are kept free. 
For the second step, the inclination and position angles are regularized to some function and only the rotation velocity and velocity dispersion are fitted.
This regularization is important to avoid that numerical oscillations in the geometrical parameters translate into nonphysical discontinuities in the rotation curve. 
We used either a Bezier curve or a constant function to regularize the geometrical angles, depending on their degree of variation with radius after step one. 
In particular, we fixed $i$ to a constant value whenever the first step returned no clear indication of a coherent warp in inclination with radius.
In these two steps, we set a $(\cos\theta)^2$ weight for the fit, because the information on the rotation velocity/warp is mostly near the major axis ($\theta\simeq0\de$), and we excluded pixels that lie within 20\de\ from the minor axis, which allowed us to minimize any effect of possible radial motions on the fit.
During the third and final step, we only fitted for the radial velocity and we fixed all other parameters to the values previously determined. 
Because radial motions have a stronger effect near the minor axis ($\theta\simeq90\de$), this time we used a $(\sin\theta)^2$ weight and we excluded pixels within 20\de\ from major axis.
Errors on best-fit parameters were estimated through BB's default Monte Carlo method \citep{DiTeodoro&Fraternali15}.

Radial velocities derived through kinematical modelling alone do not inform on whether radial motions are directed towards the center of galaxies (inflow) or towards the external regions (outflow). 
Knowing in which direction a galaxy rotates is required to interpret correctly the nature of measured radial motions. 
BB's definition of \vrad\ implies that positive (negative) velocities are outflow (inflow) motions when a galaxy is rotating clockwise, and inflow (outflow) when a galaxy is rotating anticlockwise. 
Because this information is not available in \hi\ data, we used the winding of spiral arms in optical images to determine the rotation direction, under the assumption that spiral galaxies spin with their arms trailing the direction of rotation.
Out of \galnum\ galaxies in our sample, we found that 22 galaxies are rotating clockwise and 29 anticlockwise, while it was not possible to determine the rotation direction in three S0 galaxies having no prominent spiral arms (NGC2685, NGC3941 and NGC5102). 
We used this information to correct the sign of the derived radial velocities, such that, in the remainder of the paper, $\vrad<0$ always indicates inflow and $\vrad>0$ outflow.\\

\begin{figure*}[t]
\center
\includegraphics[width=0.99\textwidth]{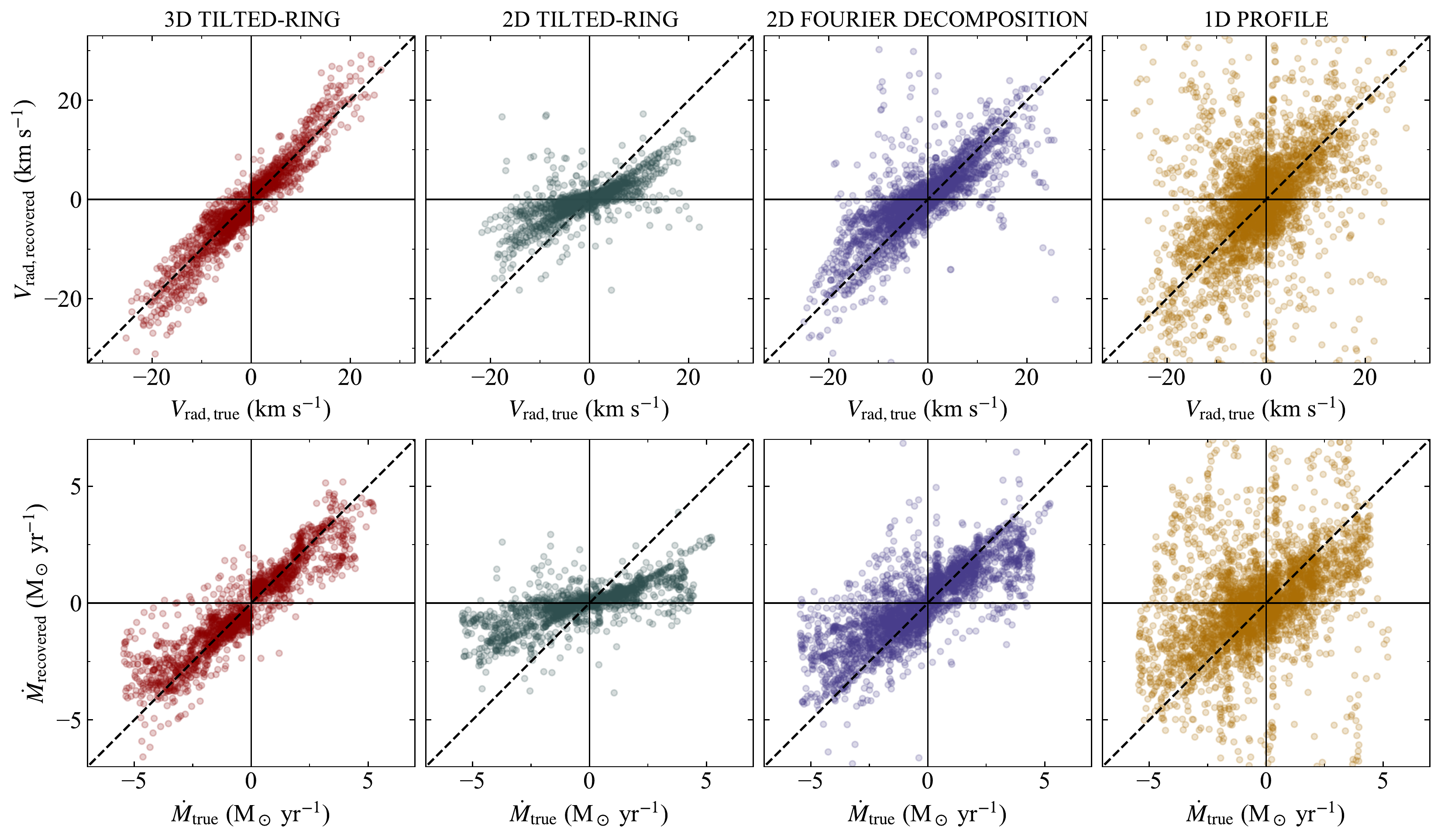}
\caption{Comparison between four different methods to derive radial velocities (top) and mass flows (bottom) in simulated disk galaxies. In red, the 3D modelling method used in this paper (Section~\ref{sec:kinmod}); in green and blue, a classical 2D tilted-ring model \citep{Begeman87} and a Fourier decomposition technique \citep{Schmidt+2016} applied to the velocity field, respectively; in gold, the average 1D profile along the minor axis.}
\label{fig:tests}
\end{figure*} 

\subsection{Surface-density and mass-flow profiles}
\label{sec:sdandmass}

Radial velocities profiles $\vrad(R)$ obtained through 3D kinematical modelling can be combined with \hi\ surface-density profiles $\Sigma_\hi(R)$ to estimate \hi-mass flow rate profiles $\dot{M}_\hi(R)$. 
We derived \hi\ intensity profiles $I(R)$ from unmasked total intensity \hi\ maps, averaging the integrated flux along elliptical rings defined by the best-fit geometrical parameters found with the kinematical modelling (\textsc{Ellprof} task in BB).
The error on $I(R)$ is calculated as the standard deviation of the integrated flux in each ring. 
Under the assumption that the \hi\ emission-line is optically thin, the intensity profile $I(R)$ can be easily converted to a face-on \hi\ mass surface-density $\Sigma_\hi(R)$ \citep[e.g.][]{Roberts+1975}:

\begin{equation}
    \label{eq:sigmaHI}
    \frac{\Sigma_\hi(R)}{[\mathrm{\mo \, pc^{-2}}]} = 8794 \frac{I(R)\cos (i(R))}{[\mathrm{Jy \,beam^{-1} \, km \, s^{-1}}]} \left( \frac{B_\mathrm{maj}B_\mathrm{min}}{[\mathrm{arcsec^2}]} \right)^{-1}
\end{equation}

\noindent where $B_\mathrm{maj}$ and $B_\mathrm{min}$ are the full widths at half maximum (FWHM) of the major and minor axes of the beam and $\cos i$ corrects for the inclination of the disk. 
The average radial flow of \hi\ mass $\mdot_\hi$ at a given radius $R$ is therefore:

\begin{equation}
    \label{eq:mdot}
    \mdot_\hi(R) = 2\pi R \, \Sigma_\hi(R) \, \vrad(R) \; \; \; .
\end{equation}

The total neutral gas mass flow is $\mdot = 1.33 \, \mdot_\hi$, where the factor 1.33 takes into account the primordial abundance of Helium. As for the radial velocity, a $\mdot<0$ indicates gas inflow (accretion) and a $\mdot>0$ indicates gas outflow.

\subsection{Tests on mock observations}
\label{sec:tests}
To validate the methodology described in previous sections, we tested it with a sample of synthetic \hi\ datacubes. 
We simulated a set of 100 galaxies, having different geometrical and kinematical properties ($\textsc{Galmod}$ task in BB). 
In particular, we explored a wide range of parameters representative of real spiral galaxies: 1) different rotation curves, including slowly-rising or steeply-rising and flat rotation curves; 2) different \hi\ surface-density profiles; 3) different warps in the outer regions of the disk, allowing a linear or quadratic variation of up to 20\de\ for the position angle and up to 10\de\ for the inclination angle (always in the range $30\de \lesssim i \lesssim 80\de$); 4) different radial inflow/outflow motions both in the inner and outer disk; 5) different disk thicknesses plus the possible presence of flares in the outer \hi\ disk.
These galaxy models were mapped into mock datacubes having observational properties, in particular spatial/spectral resolutions and levels of noise, similar to those of our \hi\ data sample.  
The main goal was to investigate to what extent a range of galaxy intrinsic features and of observational properties can hamper our ability to recover radial velocities and radial mass flow rates in disk galaxies. 

\begin{figure*}
\center
\includegraphics[width=0.983\textwidth]{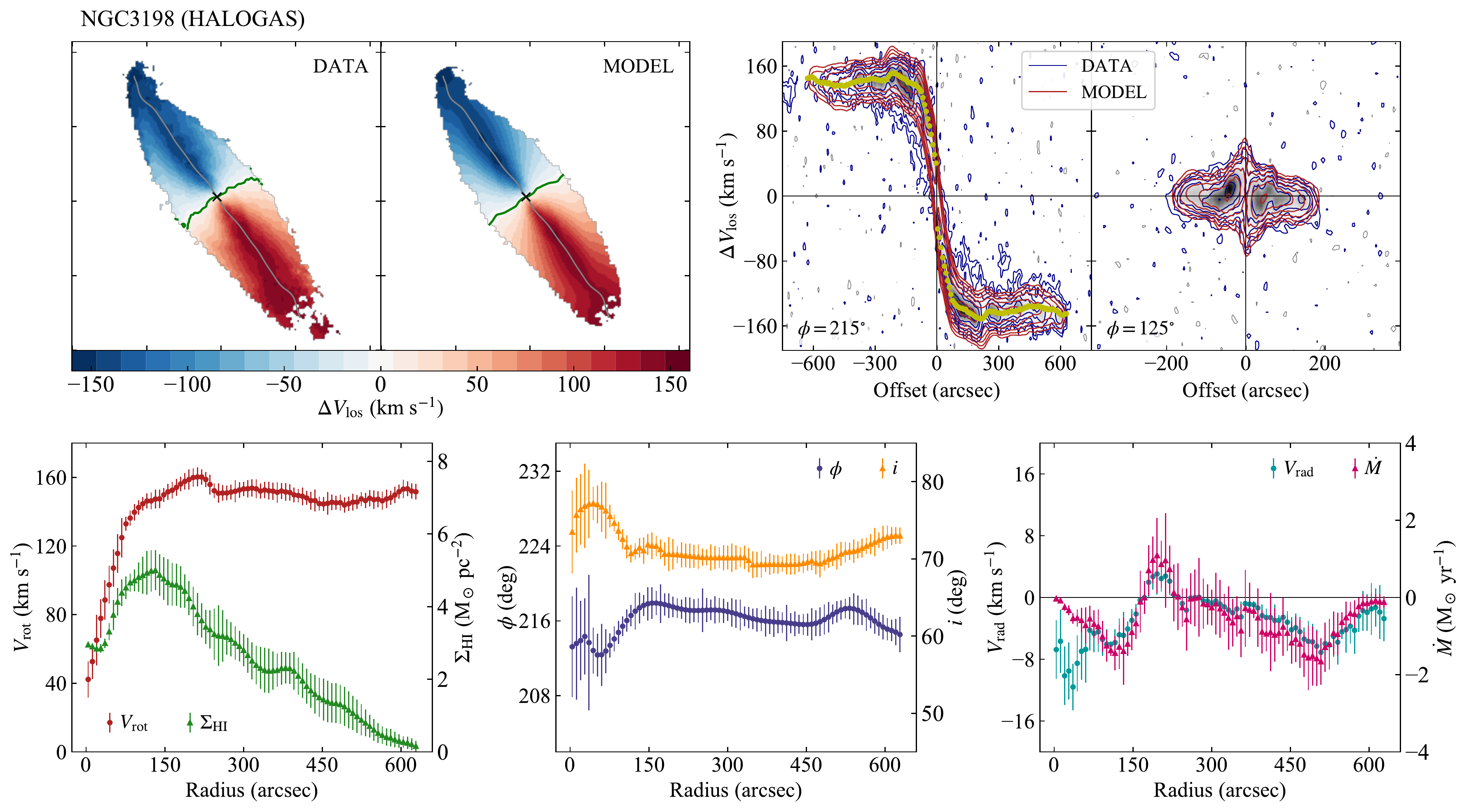}
\vspace{-5pt}
\caption{3D tilted-ring modelling of the galaxy NGC3198. \emph{Top panels}: on the left, velocity fields of the data and of the best-fit model. The black crosses denote the galaxy center, the green thick lines the systemic velocity, the grey thin lines the galaxy major axis. On the right, P-Vs slice through the average major and minor axes. Data and model are shown in blue and red contours, respectively (levels at $2^n\times\sigma_\mathrm{rms}$, with $n=0...5$). \emph{Bottom panels}: derived morpho-kinematical parameters. Left panel shows the rotation curve (red circles) and the \hi\ surface density (green triangles), mid panel the position angle (orange circles) and the inclination (blue triangles), right panel the radial velocity (cyan circles) and mass flow rate (fuchsia triangles).}
\label{fig:n3198ex}
\vspace{10pt}
\includegraphics[width=0.983\textwidth]{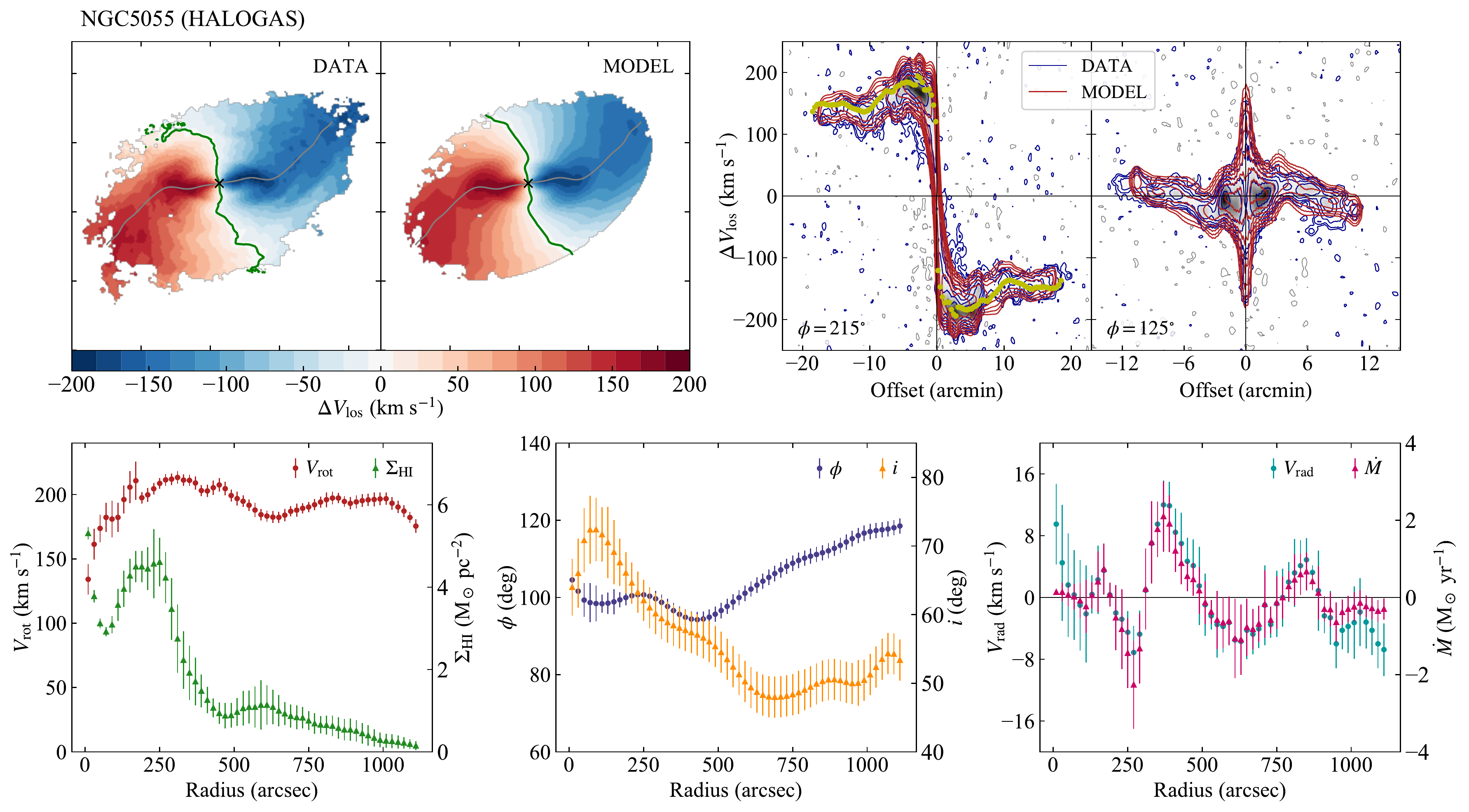}
\caption{Same as \autoref{fig:n3198ex}, but for the warped galaxy NGC5055.}
\label{fig:n5055ex}
\end{figure*} 

Our 3D kinematical modelling was performed on these 100 datacubes of simulated galaxies. 
For comparison, we also analysed these datasets and derived radial velocity profiles with a classical 2D tilted-ring model (\textsc{2dfit} task in BB), with a Fourier decomposition scheme of the velocity field (first moment) akin to that of \citet{Schmidt+2016}, and with a simple estimate of $\vrad$ through Eq.~\ref{eq:vlos} from the averaged 1D \vlos\ profile in the P-V along the minor axis.
Surface-density profiles were always derived as described in Section~\ref{sec:sdandmass}.
\autoref{fig:tests} shows plots of the ``true'' versus recovered values of $\vrad$ (top panels) and $\mdot$ (bottom) with these four methods. 
Each point in the plots denotes a measurement in a single ring.
It is clear that both a simple 2D tilted-ring modelling (green) and a 1D method (gold) are unsuitable to derive robust radial motions and mass flows, the former showing a strong bias towards lower radial flows and the latter returning very inaccurate values, mostly due the unaccounted presence of warps. 
Instead, both our 3D modelling procedure (red) and the 2D Fourier decomposition of the velocity fields (blue) recover values of $\vrad$ and $\mdot$ more consistent with the inputs. 
However, the 2D method has a larger scatter and still shows some bias towards lower radial flows, likely due to the fact that the values of $\vlos$ on a velocity field may not be representative of the motions in the disk plane in the presence of strong warps or thick disks. 
Our 3D method does not seem to produce any significant bias and recovered datapoints have a relatively small scatter of 3 $\kms$ in radial velocity and of 0.5 $\moyr$ in mass flow with respect to the one-to-one relation. This suggests that our 3D modelling procedure can be safely applied to a wide range of disk galaxies with different intrinsic and observational properties.

\section{Results}
\label{sec:results}

\subsection{Radial flows in local galaxies}

We used our 3D procedure to carefully model the kinematics of our \galnum\ galaxies. 
Each best-fit model was visually inspected and compared to the data to double-check the goodness of the fit. 
In a few difficult cases, initial parameters and options for the code were manually fine-tuned to attain a better fitting model.
\autoref{fig:n3198ex} and \autoref{fig:n5055ex} illustrate two examples of our 3D modelling applied to the galaxies NGC3198 and NGC5055 from the HALOGAS survey. 
A comparison between the data and our best-fit model is shown in the top panels by means of velocity fields (left) and P-V diagrams taken along the major and minor axes (right). 
P-Vs in particular highlight how our models (red contours) are able to reproduce almost entirely the \hi\ emission in the data (blue contours). 
Bottom panels in \autoref{fig:n3198ex} and \autoref{fig:n5055ex} show the most important model parameters derived through the modelling: the rotation curve and surface-density profile (left), the geometrical inclination and position angles (middle), the radial velocity and gas mass flow rate profile (right).
Similar plots of best-fit parameters for all galaxies in our sample can be found in \autoref{app:bestfit}.
NGC3198 is a fairly well-behaved spiral galaxy, with a regularly rotating \hi\ disk out to the outermost radii, showing no signs of a warp or of a disturbed disk ($\phi$ and $i$ are almost constant with radius). 
By contrast, NGC5055 has one of the most extreme and warped \hi\ disks in the local Universe, with variations in $\phi$ and $i$ larger than 20$\de$ and 10$\de$, respectively. 
Although these two galaxies have such different \hi\ disks, both of them only have moderate radial gas flows throughout their disk: typical radial velocities are $\mid \vrad \mid \, \lesssim 10 \, \kms$, a small fraction of the rotation velocity ($\mid \vrot/\vrad \mid \, \gtrsim 10$).
The corresponding mass flow rates are mostly $\mid \mdot \mid \, \lesssim \, 1 \, \moyr$. 
More interestingly, these radial flows do not seem to be coherent across the entire disk. 
Some regions show some clear gas inflow, for example around $R\simeq500''$ in NGC3198, while in other regions the gas flows outward, for example around $R\simeq300''$ in NGC5055 or $R\simeq150''$ in NGC3198.

Similar radial flows are detected in the majority of our \galnum\ galaxies. 
Local and global non-circular motions in a galactic disk can be induced by a variety of physical processes, like accretion of gas with different angular momentum, viscosity in the gas layer, gas streaming along a bar, spiral arms and, more in general, any perturbation of the gravitational potential \citep[e.g.,][]{Lacey+1985,Thon+1998,Armillotta+2019,Armillotta+2020,Martinez-Medina+2020}. 
In general, it is not trivial to associate unambiguously measured radial flows with one or more of these processes. However, we note that some of the largest radial motions in our sample are found in the inner regions of the disks and are sometimes associated with the presence of a prominent bar (e.g.\ NGC3351 or NGC4725) or of strong spiral arms (e.g.\ NGC3031 or NGC3992).

\begin{figure}
\hspace{-8pt}
\includegraphics[width=0.48\textwidth]{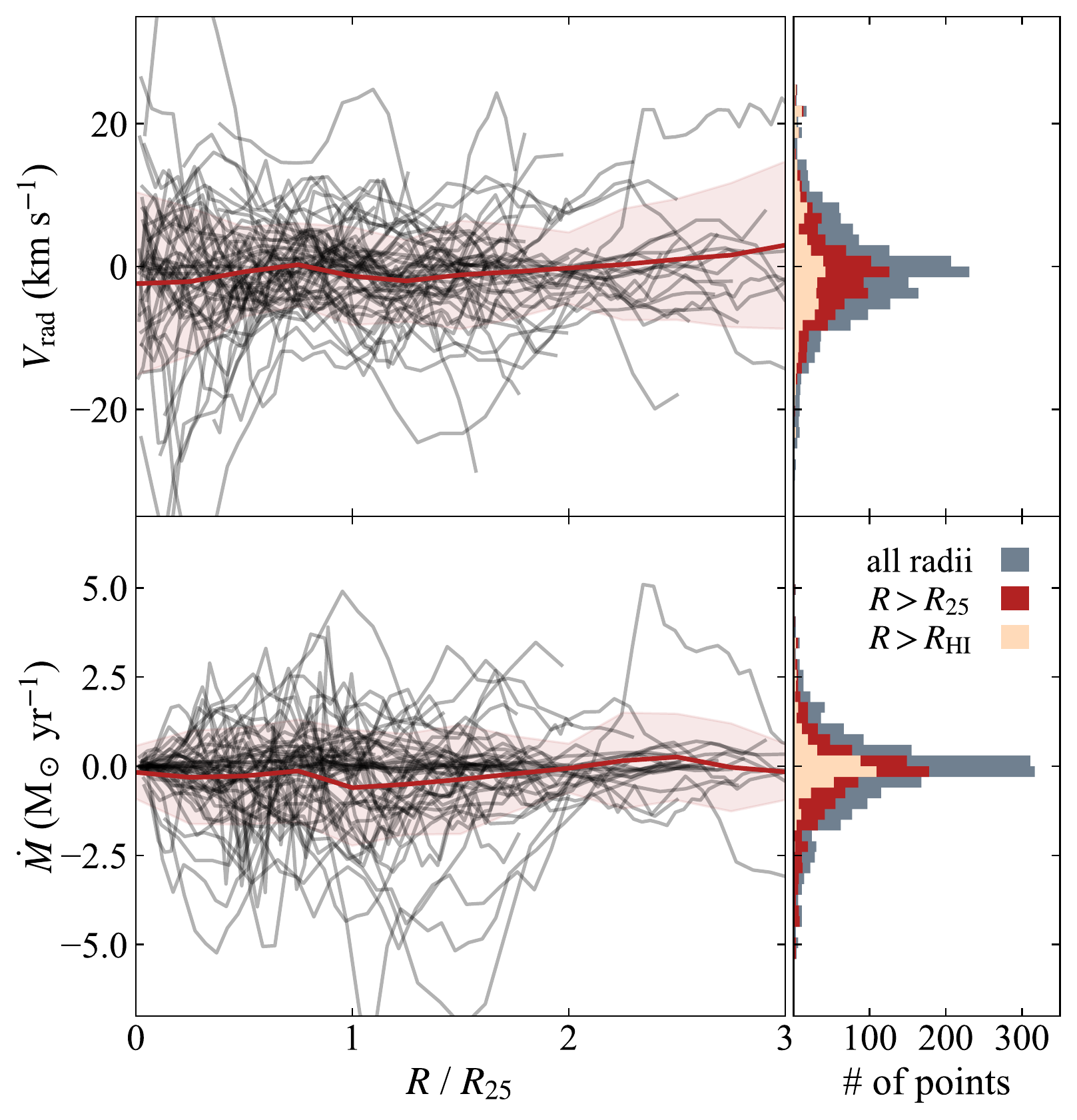}
\caption{Gas flow radial profiles for the \galnum\ analysed in this work. Radial velocities and gas mass flow rates are shown in top and bottom panels, respectively. The red thick line and shadow region denote the average and standard deviation across the entire sample. Negative (positive) values mean inflow (outflow). Histograms on the left show the corresponding distributions for all radii (blue), radii larger than the optical radius $R_{25}$ (red) and radii larger than $R_\hi$ (orange).}
\label{fig:profiles}
\end{figure}

\autoref{fig:profiles} summarizes radial velocity profiles (top) and mass flow rate profiles (bottom) for our sample. 
Radii are normalized to the radius $R_{25}$ corresponding to the 25 mag arcsec$^{-2}$ isophote in $z$-band. 
$R_{25}$ can be considered as the size of the optical disk where most star formation is occurring.
Our galaxies typically have \hi\ disks that extend up to $2-3R_{25}$.
Profiles in \autoref{fig:profiles} are akin to those of NGC3198 and NGC5055: all galaxies show some degree of radial gas flows, with radial velocities typically of a few $\kms$, but these flows do not seem to have a preferential direction.
As a consequence, the average radial velocity and mass flow rate across the sample at a given $R/R_{25}$ is nearly constant and close to zero (red thick line).
Histograms in \autoref{fig:profiles} show the frequency distributions of measured radial velocity and mass flow rate. 
We plot distributions for all radii (blue), for radii past the optical disk ($R>R_{25}$, red) and for the very outskirts of \hi\ disks (orange), i.e.\ $R>R_\hi$ where $R_\hi$ is the radius at which the \hi\ surface density drops below 1 $\mo \, \mathrm{pc^{-2}}$ ($R_\hi>R_{25}$ for all galaxies in our sample with disks more extended than $R_\hi$). 
The mean of these distributions is nearly zero (slightly negative), in both $\vrad$ and $\mdot$, with no sign of a shift towards larger flows as we move towards the outer parts of galaxy disks.
Therefore, we conclude that spiral galaxies in the local Universe do not seem to have any systematic radial gas inflow/outflow in their outer disks. 

\begin{figure*}[t]
\center
\includegraphics[width=0.95\textwidth]{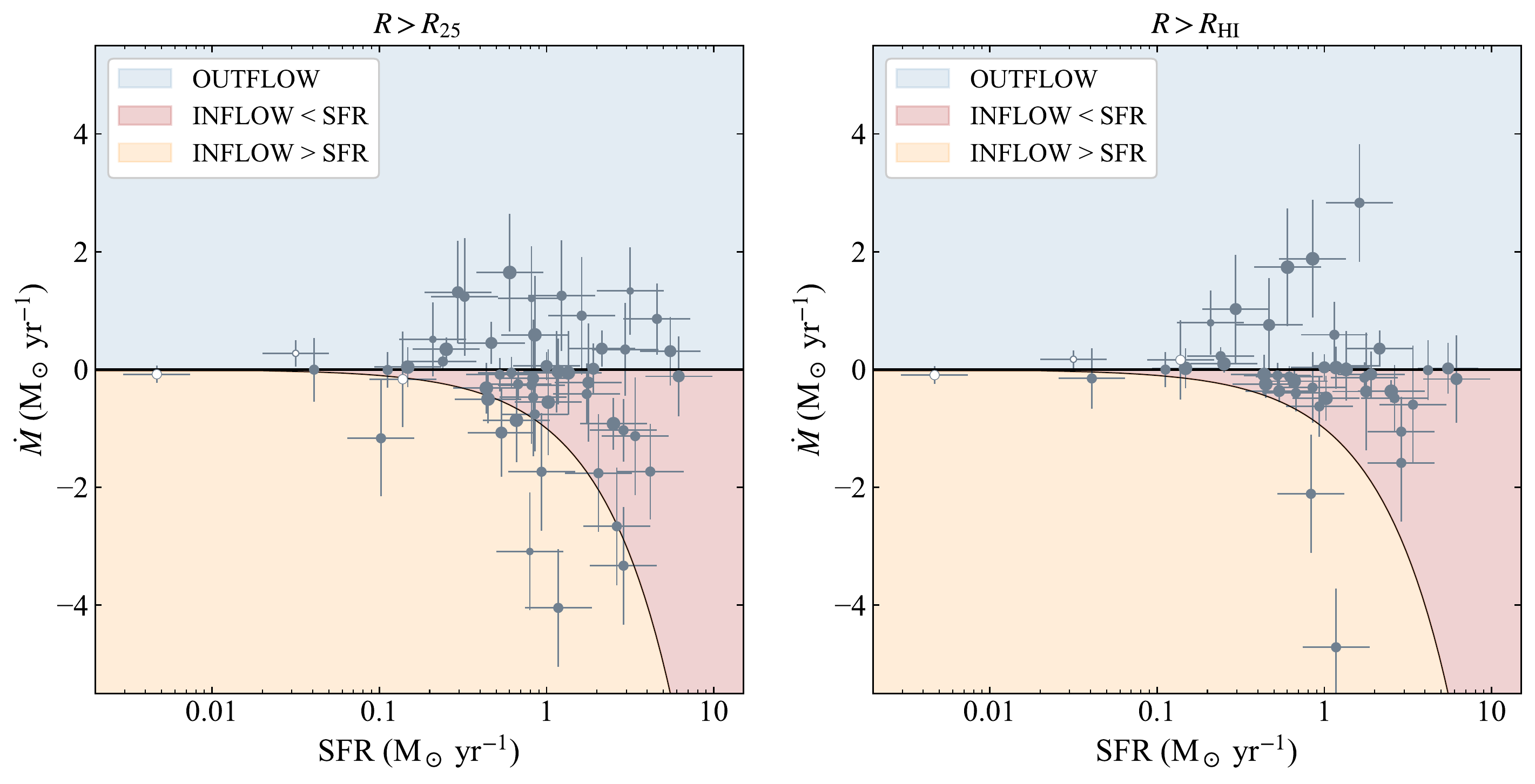}
\caption{Comparison between average mass flow rates $\mdotaver$ and star formation rates SFRs for our \galnum\ galaxies. 
The average $\mdotaver$ is calculated for $R>R_{25}$ (left) and for $R>R_{\hi}$ (right). 
We indicate in blue the region where $\mdot>0$, in red the region where $\mdot<0$ and $|\mdot| <\mathrm{SFR}$, and in yellow the region where $\mdot<0$ and $|\mdot| >\mathrm{SFR}$.
The size of points is proportional to the goodness of the best-fit kinematical model, with bigger points indicating more robust models. The three white points denote the three galaxies for which we could not infer the rotation direction and for which $\mdotaver$ might have opposite sign.}
\label{fig:acc}
\end{figure*}

\subsection{$\mdot$ vs SFR}
In \autoref{fig:acc} we compare the average mass flow rates $\mdot$ measured in this work with the SFRs of our galaxies. 
Left and right panels refer to $\mdot$ averaged over radii larger than the optical radius $R_{25}$ and over radii larger than the \hi\ radius $R_{\hi}$, respectively. 
We note that 11 galaxies in our sample that do not reach $\Sigma_\hi<1 \, \mo \mathrm{pc}^{-2}$ are removed from the right plot.
Blue regions in the plots mean outward flows. 
We highlight in red the regions where there is some inflow, but lower than the SFR, while yellow regions are where the average inflow is larger than the SFR.
About half galaxies in our sample show some inflow in their outer disks (negative $\mdot$), but for most of them the inflow rate is much smaller than the SFR. 
Eleven galaxies have inflows outside $R_{25}$ larger than the total SFR in the stellar disk, although only three at a significance level $>1\sigma$.
This number reduces to three galaxies (two at $>1\sigma$) when we average over the outermost parts of the \hi\ disk ($R>R_\hi$).
Some of the galaxies with the largest measured gas flows, either inflows or outflows, show signs of on-going or recent tidal encounters (e.g.\ NGC5033, NGC3621, NGC4651).

The mean mass flow rate over the entire sample is $-0.3\pm0.9 \moyr$ (median $-0.1 \moyr$) for $R>R_{25}$ and $-0.1\pm0.7 \moyr$ (median $-0.1 \moyr$) for $R>R_{\hi}$.
If we only consider galaxies with inflows, the mean mass flow rate is $-0.8\pm0.8 \moyr$ (median $-0.4 \moyr$) for $R>R_{25}$ and $-0.6\pm0.7 \moyr$ (median $-0.3 \moyr$) for $R>R_{\hi}$. 
These values are at a least a factor of two smaller than the average SFR of $\simeq1.4 \moyr$ of our galaxy sample.
In conclusion, although our data suggest a slight preference for inflow over outflow motions, the amount of cold gas accreted alone seems insufficient to feed the star formation occurring in the stellar disk of most local spiral galaxies. 
Average radial velocities and mass flow rates derived in this work for all galaxies can be found in \autoref{app:sample}.

\section{Discussion}
\label{sec:disc}

\subsection{Caveats}
Radial motions are notoriously difficult to measure in galaxies \citep[e.g.,][]{Wong+2004}. 
Although we used a state-of-the-art 3D technique to derive galaxy kinematics and radial motions, our approach still relies on a number of assumptions and approximations that might not be always fully appropriate for real galaxies.
First, our models are always symmetric both in the gas distribution and in the kinematics, while real \hi\ disks are known to show all sort of asymmetries, like spiral arms, asymmetric warps and morphological/kinematical lopsidedness \citep{Sancisi+2008}, especially in the outer regions. 
A few galaxies in our sample show also some tidal interactions with nearby dwarf galaxies, causing further asymmetries. 
Our method, which fits all quantities in symmetric rings, tends to smooth out these asymmetries and to find average values that can reproduce the bulk of the gas in a ring. 

A second limitation comes from the assumption that gas always move in circular orbits. 
This approximation holds for the inner regions of \hi\ disks, but it probably breaks for warped orbits as they move away from the midplane. This would affect both the rotation velocity and the radial velocity that we estimated with our 3D modelling. 
Finally, we also assumed that \hi\ disks have a constant scale height and that there are no significant vertical motions. However, \hi\ disks are expected to flare in the outermost regions \citep[e.g.,][]{Kalberla+2008}, where the gravitational potential of the disk becomes weaker. 
Some vertical motions, likely of a magnitude similar to the measured radial motions, are also expected because, for example, of star formation feedback in the disk. 
Parameters that might be affected by an accounted presence of flares and vertical motions include the galaxy inclination and the rotation/radial velocity.

Given the above, the magnitude of radial inflow/outflow measured in this paper for individual galaxies should be regarded as significantly uncertain. 
However, we do not expect the above assumptions to produce a bias in the measured radial flows over a statistical sample, but only to increase the scatter of our measurements. 
For example, if all galaxies in our sample were experiencing systematic and strong radial inflows in their outermost regions, this would produce a detectable effect when averaging over the entire population, despite all the approximations in our modelling approach.

\subsection{Comparison with previous studies}
Only few former works have tried to quantify radial gas flows in \hi\ data of local galaxies \citep{Wong+2004,Trachternach+2008,Schmidt+2016}. 
While we performed a full 3D modelling of \hi\ data cubes, all previous studies derived radial motions through a harmonic decomposition of the 2D velocity field (see blue dots in \autoref{fig:tests}).
\citet{Wong+2004} attempted the first systematic investigation of radial gas flows in 7 nearby spirals, three of which are included in our sample as well (NGCs 4414, 5033 and 5055), mostly focusing on the inner regions of the disks. 
Using both \hi\ and CO data, they concluded that there is no
unequivocal evidence for significant radial inflows and that most non-circular motions are attributable to the presence of bars and spiral structures. 
\citet{Trachternach+2008} studied the effect of non-circular motions on the dark matter halo by analysing a sample of 19 galaxies from the THINGS survey, including both spiral and dwarf galaxies (12 in common with our study).
They also found that, on average, non-circular motions are smaller than $\sim 10 \, \kms$ and only a small fraction of the rotation velocity ($\sim 5\%$).
Therefore, our findings on a larger sample of \galnum\ galaxies are in good agreement with both \citet{Wong+2004} and \citet{Trachternach+2008}, i.e.\ we confirm that, in general, significant radial motions are not detected in local spiral galaxies.

The most recent observational study on radial gas flows is that of \citet{Schmidt+2016}.
They re-analysed 10 galaxies from the THINGS survey using a more sophisticated Fourier decomposition technique of the velocity field and compared the derived mass flow rates with star formation rate profiles. 
Nine of these galaxies are also included in our sample (all but NGC925). 
\citet{Schmidt+2016} claimed that, in at least 5 out of 10 galaxies (NGCs 2403, 3198, 2903, 6946 and 7331), average mass inflow rates outside the optical disks are quite larger than the star formation rates. 
For the remaining 5 galaxies, the picture looks more complicated, showing either inflow, outward motions or complex kinematic signatures.
We note that the shape of radial profiles showed in \citet{Schmidt+2016} are overall in good agreement with ours (see \autoref{app:bestfit}), and the magnitude of their derived radial velocities are also generally within a few $\kms$ from those measured in our work. 
However, their mass flow rates are consistently larger than ours. 
We argue that this discrepancy might be ascribable to the different masking used for the derivation of \hi\ column densities. 
While we used unmasked total \hi\ maps, \citet{Schmidt+2016} relied on THINGS 0th-moment maps, obtained with a quite aggressive masking \citep{Walter+2008}.
According to our tests with mock data, this type of masking may cause an overestimation of the \hi\ column density in the faint outskirts of disks, because noisy pixels with positive values may be included in the mask without their negative counterparts. In particular, we found that this effect can be important with THINGS data, where the properties of the noise are strongly uneven across the field of view.

\subsection{No significant radial accretion?}

Theories of galaxy evolution predict that cosmological gas is firstly accreted onto galaxies at large radii and then radially moves towards the inner disk to feed the star-formation process \citep{Stewart+2013,Ho+2019}.
A few recent works have studied radial gas accretion in Milky-Way-like galaxies in cosmological hydrodynamical simulation \citep{Nuza+2019,Okalidis+2021,Trapp+2021}. 
These studies found that the expected gas inflow radial motions in most star-forming galaxies are of the order of a few $\kms$ within the inner optical disk and can reach $10-15 \; \kms$ in the outskirts.
Predicted mass flows are of the order of a few $\moyr$, usually sufficient to sustain the star formation for long times. 

From our analysis, a more complex picture arises. 
First of all, only half of the galaxies in our sample show some radial gas inflows, suggesting that radial accretion is present in only a fraction of star-forming disks.
For most of these galaxies, we measured radial velocities of a few up to about $10$ $\kms$, in agreement with expectations from simulations, but without observing the predicted increment in the outer regions of the \hi\ disks. 
Moreover, we found that radial motions are highly variable and can easily transition from inflow to outflow throughout the disk, likely due to local perturbing structures like spiral arms or bars.
Although the measured amount of gas radially accreted can contribute to sustain the current star formation rate in some galaxies, it seems insufficient to support it entirely in most of them. 
In fact, only $2-3$ galaxies in our sample have $\mdot>\mathrm{SFR}$ at a $>1\sigma$ level, while most galaxies seem to have little or no radial accretion.
However, we also stress that the current SFR of a galaxy and the current gas inflow rate may not be easily related to each other, as gas accretion and star formation act on different timescales.
Observed gas inflowing from the external regions of the disk with typical velocities of a few $\kms$ will be available for star formation in the inner regions on timescales $\gtrsim 1$ Gyr. 
Star formation and accretion rates are directly comparable only if we assume a quasi steady-state evolution over the past few Gyr.

Our study suggests that the observed amount of gas radially inflowing from the intergalactic and circumgalactic media at the edges of most local spiral galaxies will not be enough to sustain star formation at its current rate in the near future.
This implies that either the SFR in most spiral galaxies will be declining over time or that other dominant forms of gas accretion, not requiring significant radial inflows, might be present. 
For example, some gas might be falling down nearly vertically onto the inner star forming disk. 
We know that both the Milky Way \citep[e.g.][]{Wakker+1997,Putman+2012}, M31 and M33 \citep[e.g.,][]{Westmeier+2005,Kerp+2016} are surrounded by several massive \hi\ structures (``high-velocity clouds'') residing within their inner halos, which is likely a common feature of most spiral galaxies. 
However, estimates of accretion rates from these \hi\ structures are relatively small \citep{Sancisi+2008,Putman+2012}.
A significant reservoir of fresh cold/warm gas might come from the extended and diffuse CGM observed in absorption out to $\sim150$ kpc in many star-forming galaxies \citep[][]{Tumlinson+2017}, but the mode of a possible accretion of this gas is still poorly understood. 
Another intriguing scenario, compatible with no significant radial inflow motions, proposes that most accretion might happen at the disk-halo interface through the condensation of the hot corona triggered by supernova feedback \citep{Marinacci+2010,Armillotta+2016}. 
All these mechanisms may be working together to bring enough fresh gas into star-forming disks.
Finally, another possibility is that radial accretion is an episodic and fast rather than a continuous process, triggered for example by tidal encounters and/or disk instabilities.  
In any case, our results seem to discourage a dominant, secular, radial accretion scenario, where most fresh gas from the CGM/IGM is steadily transported inwards from the outermost to the innermost regions of disks.

\section{Summary and conclusions}
\label{sec:concl}
In this work, we used the best \hi\ data currently available to determine radial motions and radial gas flows in local spiral galaxies.
We developed a multi-step 3D technique to break the degeneracy between warps and radial motions and we derived accurate kinematical models and \hi\ surface-density profiles for 54 nearby spiral galaxies.
For most galaxies, we found radial velocities, directed either inwards (inflow) or outwards (outflow), of only a few $\kms$, without any clear increase moving towards the outermost regions of \hi\ disks.
These radial velocities translate into cold gas mass inflow/outflow rates of only a few tenths of $\moyr$ in most cases.
The largest radial flows are generally associated with structures like bars and spiral arms in the inner disks, and with tidal interactions and/or minor mergers in the outer disks.
When averaging over the entire sample, inflow motions seem slightly favoured over outflow motions, but, in general, we did not find any kinematical signature for systematic radial inflows in spiral galaxies in the nearby Universe.

Instantaneous star formation rates of our galaxies were compared with the average gas mass flow rates outside the optical radius ($R>R_{25}$) and in the outermost portions of the \hi\ disks ($R>R_\hi$). 
About half of the galaxies in our sample show some degree of inflow, but only a few of them have measured mass inflow rates that could significantly sustain the star formation process in the inner disk. 
We calculated a mean mass flow rate for the entire sample of $-0.3\pm0.9 \moyr$ for $R>R_{25}$ and of $-0.1\pm0.7 \moyr$ for $R>R_{\hi}$.
These mass flow rates are several times smaller than the average star formation rate of $1.4 \moyr$.
We conclude that, despite the uncertainties in our measurements, we do not see any convincing evidence for a secular radial gas accretion process from the IGM/CGM onto star-forming galaxies in the local Universe.
In most galaxies, additional channels of accretion that do not require a significant radial transport of mass in the gaseous disk are needed to feed and sustain the star formation process over the next few Gyrs.

\vspace*{0.5cm}
\begin{acknowledgments}
EDT thanks F.\ Fraternali for fruitful discussions on the topics of this paper. 
EDT was supported by the US National Science Foundation under grant 1616177.
\end{acknowledgments}

\facilities{VLA, WSRT, ATCA}

\software{\bba\ \citep{DiTeodoro&Fraternali15,DiTeodoroBB}, CASA \citep{Raba+2020}, \textsc{Astropy} \citep{astropy:2018}}

\bibliography{biblio_rm}{}
\bibliographystyle{aasjournal}

\newpage 
\appendix
\section{Properties of galaxy sample}
\label{app:sample}
This Appendix contains a table with the sample of \galnum\ galaxies analysed in this work and their derived properties. Columns are as follow: (1) Galaxy name.(2) Distances from EDD database \citep{Tully+2016}. (3) Inclination angle derived with kinematic modelling. (4) Star formation rates taken from \citet{Leroy+2019} or calculated from WISE $W3$-band luminosities. (5) Direction of rotation: A = anticlockwise, C = clockwise, ? = undefined. (6) Average radial velocity for $R>R_{25}$. Negative/positive values indicate inflow/outflow. (7) Average radial velocity for $R>R_\hi$. Eleven galaxies do not reach $R>R_\hi$. (8) Average gas mass flow rate for $R>R_{25}$.  Negative/positive values indicate inflow/outflow. (9) Average gas mass flow rate for $R>R_\hi$. (10) \hi\ datacubes from: 1 = THINGS \citep{Walter+2008}, 2 = HALOGAS \citep{Heald+2011}, 3 = VIVA \citep{Chung+2009}, 4 = WHISP \citep{vanderHulst+2001}, 5 = \citet{Richards+2016}, 6 = LVHIS \citep{Koribalski+2018}, 7 = HIX \citep{Lutz+2017}, 8 = ATLAS$^{\mathrm{3D}}$ \hi\ \citep{Serra+2012}, 9 = \citet{Braun+2009}.

\LTcapwidth=\textwidth

\begin{longtable*}{llcccccccc}
\caption{Galaxy sample analysed in this work.  } \label{tab:sample}\\
\noalign{\vspace{5pt}}\hline\hline\noalign{\vspace{5pt}}
Name & $D$ & $i$ & SFR & RD & $\langle \vrad_{,25} \rangle$ & $\langle \vrad_{,\hi} \rangle$ & $\langle \mdot_{25} \rangle$ & $\langle \mdot_{\rm HI} \rangle$ & Ref. \\
 & Mpc & deg & $\moyr$ &  & $\kms$ & $\kms$ & $\moyr$ & $\moyr$ & \\
(1)  & (2) & (3) & (4) & (5) & (6) & (7) & (8) & (9) & (10)\\
\noalign{\vspace{5pt}}\hline\noalign{\vspace{5pt}}
ESO097-G013	&	$4.2\pm0.7$	&	55	&	$5.50\pm2.28$	&	A	&	$+2\pm3$	&	$+0\pm3$	&	$+0.31\pm0.58$	&	$+0.02\pm0.43$	&	6	\\
NGC0210	&	$20.9\pm2.7$	&	52	&	$0.32\pm0.15$	&	C	&	$+4\pm4$	&	-	&	$+1.24\pm1.48$	&	-	&	8	\\
NGC0224	&	$0.77\pm0.04$	&	77	&	$2.51\pm1.16$	&	A	&	$-6\pm3$	&	$-7\pm4$	&	$-0.92\pm0.44$	&	$-0.37\pm0.19$	&	9	\\
NGC0253	&	$3.6\pm0.3$	&	80	&	$4.57\pm2.11$	&	C	&	$+5\pm3$	&	-	&	$+0.86\pm0.60$	&	-	&	6	\\
NGC0289	&	$23.3\pm3.2$	&	45	&	$1.17\pm0.54$	&	A	&	$-4\pm3$	&	$-16\pm4$	&	$-4.05\pm1.50$	&	$-4.71\pm1.50$	&	7	\\
NGC0949	&	$10.3\pm1.3$	&	68	&	$0.24\pm0.11$	&	C	&	$+4\pm3$	&	$+7\pm3$	&	$+0.14\pm0.15$	&	$+0.23\pm0.15$	&	2	\\
NGC1003	&	$9.9\pm2.7$	&	79	&	$0.21\pm0.10$	&	A	&	$+4\pm3$	&	$+7\pm3$	&	$+0.51\pm0.63$	&	$+0.79\pm0.55$	&	2	\\
NGC1313	&	$4.2\pm0.4$	&	44	&	$0.68\pm0.31$	&	C	&	$-8\pm3$	&	$-13\pm4$	&	$-0.25\pm0.25$	&	$-0.40\pm0.31$	&	6	\\
NGC2403	&	$3.2\pm0.3$	&	60	&	$0.45\pm0.21$	&	C	&	$-3\pm3$	&	$-3\pm3$	&	$-0.50\pm0.41$	&	$-0.25\pm0.24$	&	2	\\
NGC2685	&	$37.8\pm3.9$	&	61	&	$0.14\pm0.06$	&	?	&	$+0\pm3$	&	$+2\pm5$	&	$-0.16\pm0.81$	&	$+0.16\pm0.67$	&	8	\\
NGC2841	&	$14.1\pm1.9$	&	71	&	$0.85\pm0.39$	&	A	&	$+9\pm3$	&	$+21\pm3$	&	$+0.59\pm1.25$	&	$+1.88\pm1.50$	&	1	\\
NGC2903	&	$8.9\pm1.3$	&	66	&	$0.44\pm0.20$	&	A	&	$-3\pm3$	&	$-3\pm3$	&	$-0.32\pm0.43$	&	$-0.08\pm0.34$	&	1	\\
NGC2985	&	$16.6\pm3.6$	&	50	&	$0.81\pm0.37$	&	C	&	$-1\pm3$	&	-	&	$-0.26\pm1.04$	&	-	&	4	\\
NGC3001	&	$26.7\pm7.4$	&	53	&	$1.78\pm0.82$	&	A	&	$-3\pm4$	&	$-6\pm4$	&	$-0.22\pm1.15$	&	$-0.37\pm1.21$	&	7	\\
NGC3031	&	$3.7\pm0.3$	&	63	&	$0.54\pm0.25$	&	A	&	$-5\pm3$	&	$-22\pm5$	&	$-1.07\pm0.75$	&	$-0.36\pm0.19$	&	1	\\
NGC3184	&	$13.0\pm1.6$	&	24	&	$1.35\pm0.62$	&	C	&	$+0\pm3$	&	$+2\pm4$	&	$-0.18\pm0.71$	&	$+0.11\pm0.35$	&	1	\\
NGC3198	&	$13.8\pm1.1$	&	73	&	$1.02\pm0.47$	&	A	&	$-2\pm3$	&	$-4\pm4$	&	$-0.55\pm0.90$	&	$-0.49\pm0.38$	&	2	\\
NGC3344	&	$10.0\pm1.3$	&	29	&	$0.83\pm0.38$	&	C	&	$+0\pm3$	&	-	&	$-0.15\pm1.10$	&	-	&	4	\\
NGC3351	&	$10.5\pm0.8$	&	45	&	$1.17\pm0.54$	&	A	&	$+0\pm3$	&	$+1\pm4$	&	$-0.03\pm0.56$	&	$+0.03\pm0.36$	&	1	\\
NGC3486	&	$12.7\pm2.8$	&	45	&	$1.15\pm0.53$	&	A	&	$+1\pm3$	&	$+6\pm5$	&	$-0.04\pm0.69$	&	$+0.59\pm0.55$	&	5	\\
NGC3521	&	$13.2\pm3.1$	&	70	&	$2.63\pm1.21$	&	A	&	$-7\pm3$	&	$-6\pm3$	&	$-2.66\pm1.35$	&	$-0.49\pm0.48$	&	1	\\
NGC3556	&	$9.6\pm1.3$	&	78	&	$1.23\pm0.57$	&	C	&	$+6\pm4$	&	-	&	$+1.25\pm0.94$	&	-	&	4	\\
NGC3621	&	$6.7\pm0.5$	&	65	&	$0.83\pm0.38$	&	C	&	$-3\pm3$	&	$-17\pm5$	&	$-0.47\pm1.41$	&	$-2.11\pm1.50$	&	1	\\
NGC3627	&	$10.6\pm1.0$	&	64	&	$3.16\pm1.46$	&	A	&	$+9\pm4$	&	-	&	$+1.33\pm0.74$	&	-	&	1	\\
NGC3675	&	$14.4\pm3.1$	&	67	&	$1.00\pm0.46$	&	A	&	$+1\pm4$	&	$+1\pm3$	&	$+0.06\pm0.24$	&	$+0.04\pm0.23$	&	5	\\
NGC3898	&	$22.1\pm6.1$	&	66	&	$0.30\pm0.14$	&	A	&	$+8\pm3$	&	$+11\pm3$	&	$+1.31\pm0.87$	&	$+1.03\pm0.92$	&	4	\\
NGC3941	&	$12.2\pm1.7$	&	50	&	$0.03\pm0.01$	&	?	&	$+5\pm3$	&	$+6\pm4$	&	$+0.28\pm0.23$	&	$+0.17\pm0.15$	&	5	\\
NGC3992	&	$17.1\pm4.7$	&	62	&	$0.85\pm0.39$	&	A	&	$-4\pm3$	&	$+0\pm4$	&	$-0.76\pm0.63$	&	$-0.31\pm0.60$	&	5	\\
NGC4192	&	$16.8\pm3.7$	&	77	&	$1.35\pm0.62$	&	A	&	$+0\pm3$	&	$+0\pm4$	&	$-0.01\pm0.51$	&	$+0.06\pm0.59$	&	3	\\
NGC4258	&	$7.6\pm0.6$	&	67	&	$0.93\pm0.43$	&	A	&	$-6\pm3$	&	$-6\pm4$	&	$-1.74\pm1.08$	&	$-0.63\pm0.52$	&	2	\\
NGC4414	&	$17.7\pm1.8$	&	56	&	$2.88\pm1.33$	&	A	&	$-11\pm3$	&	$-12\pm4$	&	$-1.03\pm0.53$	&	$-1.05\pm0.59$	&	2	\\
NGC4535	&	$15.8\pm1.3$	&	36	&	$2.04\pm0.94$	&	A	&	$-7\pm4$	&	-	&	$-1.76\pm1.22$	&	-	&	3	\\
NGC4536	&	$15.2\pm1.9$	&	66	&	$2.95\pm1.36$	&	C	&	$+1\pm3$	&	-	&	$+0.34\pm0.79$	&	-	&	3	\\
NGC4548	&	$16.2\pm2.2$	&	44	&	$0.52\pm0.24$	&	C	&	$-2\pm5$	&	$-3\pm4$	&	$-0.08\pm0.29$	&	$-0.10\pm0.21$	&	3	\\
NGC4559	&	$8.9\pm0.8$	&	66	&	$0.66\pm0.30$	&	A	&	$-4\pm4$	&	$-2\pm3$	&	$-0.86\pm0.71$	&	$-0.20\pm0.25$	&	2	\\
NGC4651	&	$16.8\pm4.6$	&	51	&	$0.79\pm0.37$	&	A	&	$-12\pm3$	&	-	&	$-3.09\pm1.50$	&	-	&	3	\\
NGC4698	&	$19.6\pm4.2$	&	70	&	$0.15\pm0.07$	&	C	&	$+0\pm3$	&	$+1\pm4$	&	$+0.04\pm0.34$	&	$+0.02\pm0.34$	&	3	\\
NGC4725	&	$12.4\pm1.1$	&	56	&	$0.81\pm0.37$	&	C	&	$+7\pm3$	&	-	&	$+1.21\pm0.89$	&	-	&	4	\\
NGC4736	&	$4.4\pm0.4$	&	40	&	$0.47\pm0.22$	&	C	&	$+10\pm3$	&	$+19\pm3$	&	$+0.45\pm0.36$	&	$+0.76\pm0.80$	&	1	\\
NGC5005	&	$18.4\pm2.1$	&	69	&	$2.14\pm0.99$	&	A	&	$+7\pm3$	&	$+7\pm3$	&	$+0.35\pm0.31$	&	$+0.35\pm0.31$	&	5	\\
NGC5033	&	$19.0\pm2.7$	&	66	&	$2.88\pm1.33$	&	A	&	$-7\pm4$	&	$-7\pm3$	&	$-3.33\pm1.50$	&	$-1.58\pm1.13$	&	5	\\
NGC5055	&	$8.9\pm0.7$	&	55	&	$1.91\pm0.88$	&	C	&	$+0\pm3$	&	$-1\pm3$	&	$+0.01\pm0.52$	&	$-0.08\pm0.43$	&	2	\\
NGC5102	&	$3.8\pm0.3$	&	71	&	$0.01\pm0.01$	&	?	&	$-4\pm2$	&	$-5\pm2$	&	$-0.08\pm0.15$	&	$-0.09\pm0.15$	&	6	\\
NGC5161	&	$25.5\pm5.1$	&	69	&	$1.62\pm0.75$	&	A	&	$+5\pm3$	&	$+16\pm3$	&	$+0.91\pm1.50$	&	$+2.83\pm1.50$	&	7	\\
NGC5236	&	$4.9\pm0.5$	&	40	&	$4.17\pm1.92$	&	C	&	$-7\pm3$	&	$-1\pm3$	&	$-1.73\pm0.81$	&	$-0.01\pm0.51$	&	6	\\
NGC5350	&	$30.9\pm5.5$	&	54	&	$1.74\pm0.80$	&	C	&	$-2\pm3$	&	$-1\pm3$	&	$-0.42\pm0.52$	&	$-0.14\pm0.26$	&	8	\\
NGC5582	&	$28.4\pm3.9$	&	53	&	$0.04\pm0.02$	&	C	&	$-1\pm3$	&	$+0\pm4$	&	$-0.01\pm0.54$	&	$-0.15\pm0.52$	&	8	\\
NGC6798	&	$39.3\pm10.5$	&	62	&	$0.11\pm0.05$	&	A	&	$+0\pm3$	&	$+0\pm3$	&	$-0.01\pm0.30$	&	$-0.00\pm0.30$	&	8	\\
NGC6946	&	$7.7\pm1.2$	&	25	&	$6.17\pm2.84$	&	C	&	$-1\pm3$	&	$-2\pm3$	&	$-0.11\pm0.68$	&	$-0.16\pm0.75$	&	1	\\
NGC7217	&	$15.0\pm4.1$	&	34	&	$0.62\pm0.28$	&	C	&	$-1\pm3$	&	$-3\pm4$	&	$-0.05\pm0.27$	&	$-0.12\pm0.18$	&	4	\\
NGC7331	&	$14.7\pm1.1$	&	77	&	$3.39\pm1.56$	&	C	&	$-5\pm3$	&	$-5\pm3$	&	$-1.13\pm1.50$	&	$-0.60\pm1.50$	&	1	\\
NGC7793	&	$3.6\pm0.3$	&	46	&	$0.25\pm0.12$	&	A	&	$+5\pm3$	&	$+4\pm3$	&	$+0.34\pm0.20$	&	$+0.10\pm0.15$	&	1	\\
UGC02080	&	$14.4\pm3.1$	&	23	&	$0.10\pm0.05$	&	A	&	$-5\pm3$	&	-	&	$-1.17\pm0.98$	&	-	&	4	\\
UGC02953	&	$21.5\pm3.9$	&	49	&	$0.60\pm0.28$	&	A	&	$+8\pm4$	&	$+9\pm4$	&	$+1.65\pm1.39$	&	$+1.74\pm1.42$	&	4	\\

\noalign{\vspace{2pt}}\hline
\noalign{\vspace{2pt}}

\end{longtable*}

\newpage
\section{Best-fit parameters}
\label{app:bestfit}
In this Appendix, we show plots of derived morpho-kinematical parameters for all \galnum\ galaxies in our sample. Different rows indicate different galaxies. For each galaxy, we plot the rotation curve (red circles) and the \hi\ surface density (green triangles) in the left panel, the position angle (orange circles) and the inclination (blue triangles) in the middle panel, the radial velocity (cyan circles) and mass flow rate (fuchsia triangles) in the right panel.

\begin{figure}
	\includegraphics[width=.99\textwidth]{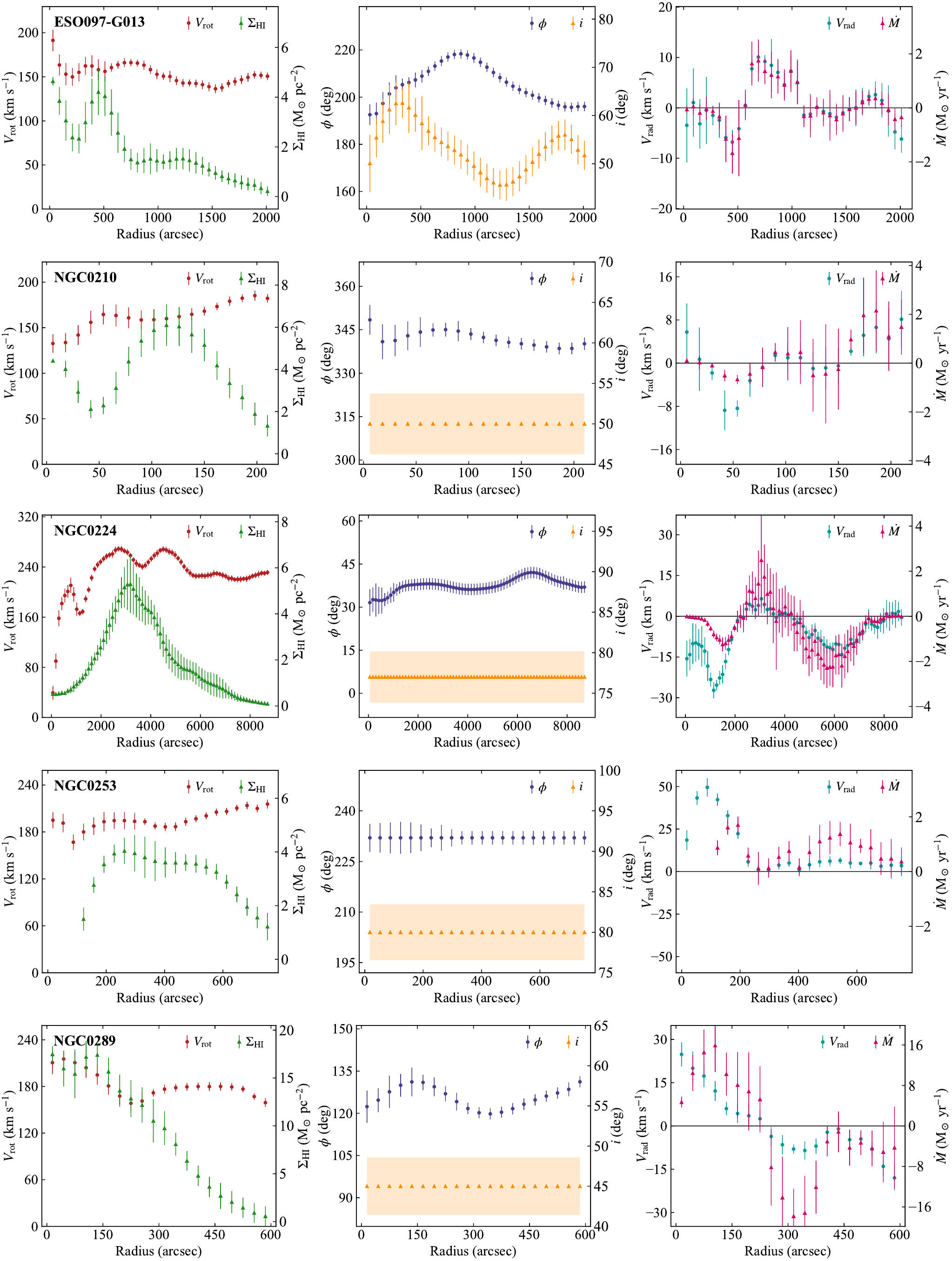}
    \caption{Kinematical parameters derived through 3D modelling.}
    \label{fig:appfig}
\end{figure}
\begin{figure*}
	\includegraphics[width=.99\textwidth]{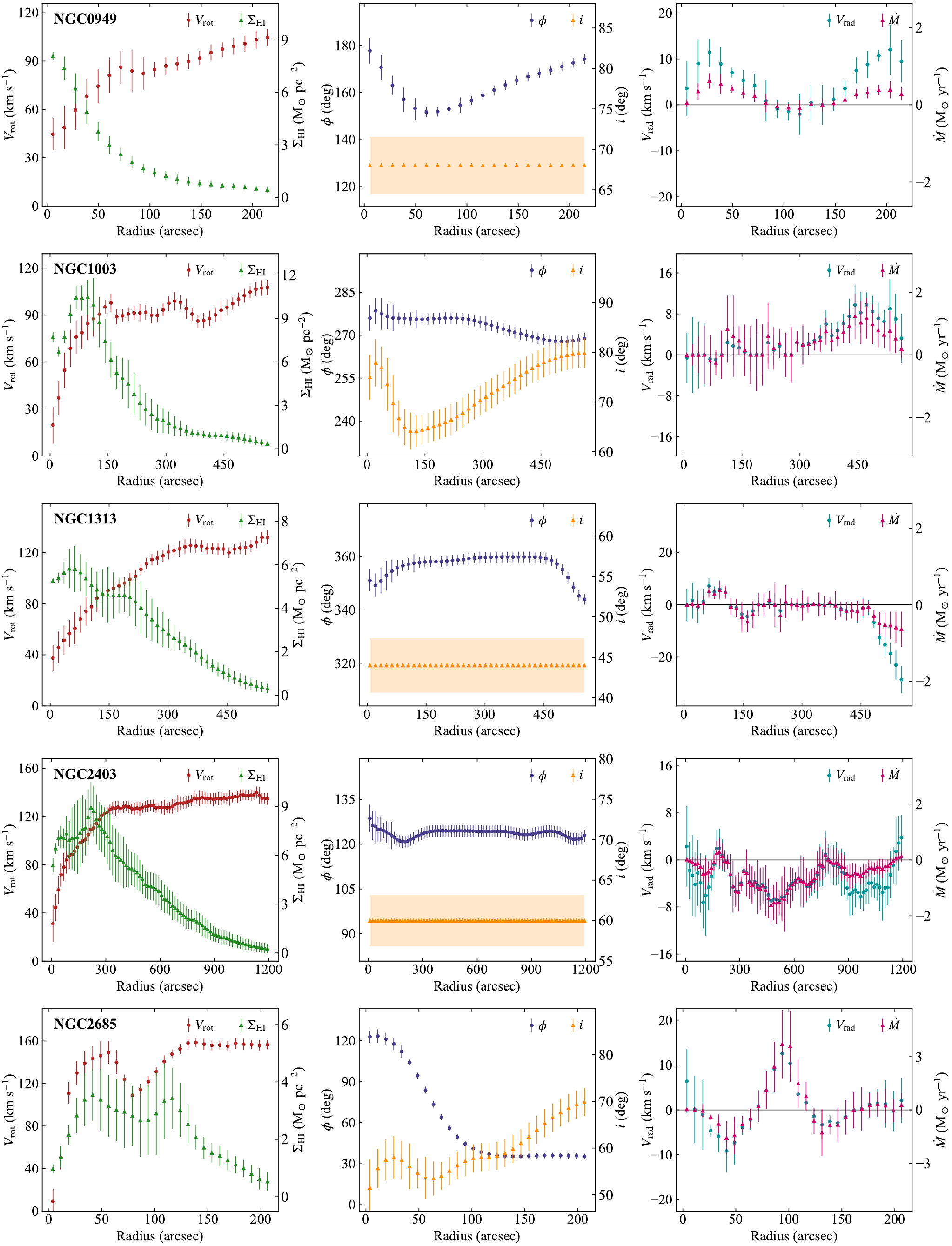}
    \caption{Continued}
\end{figure*}
\begin{figure*}
	\includegraphics[width=.99\textwidth]{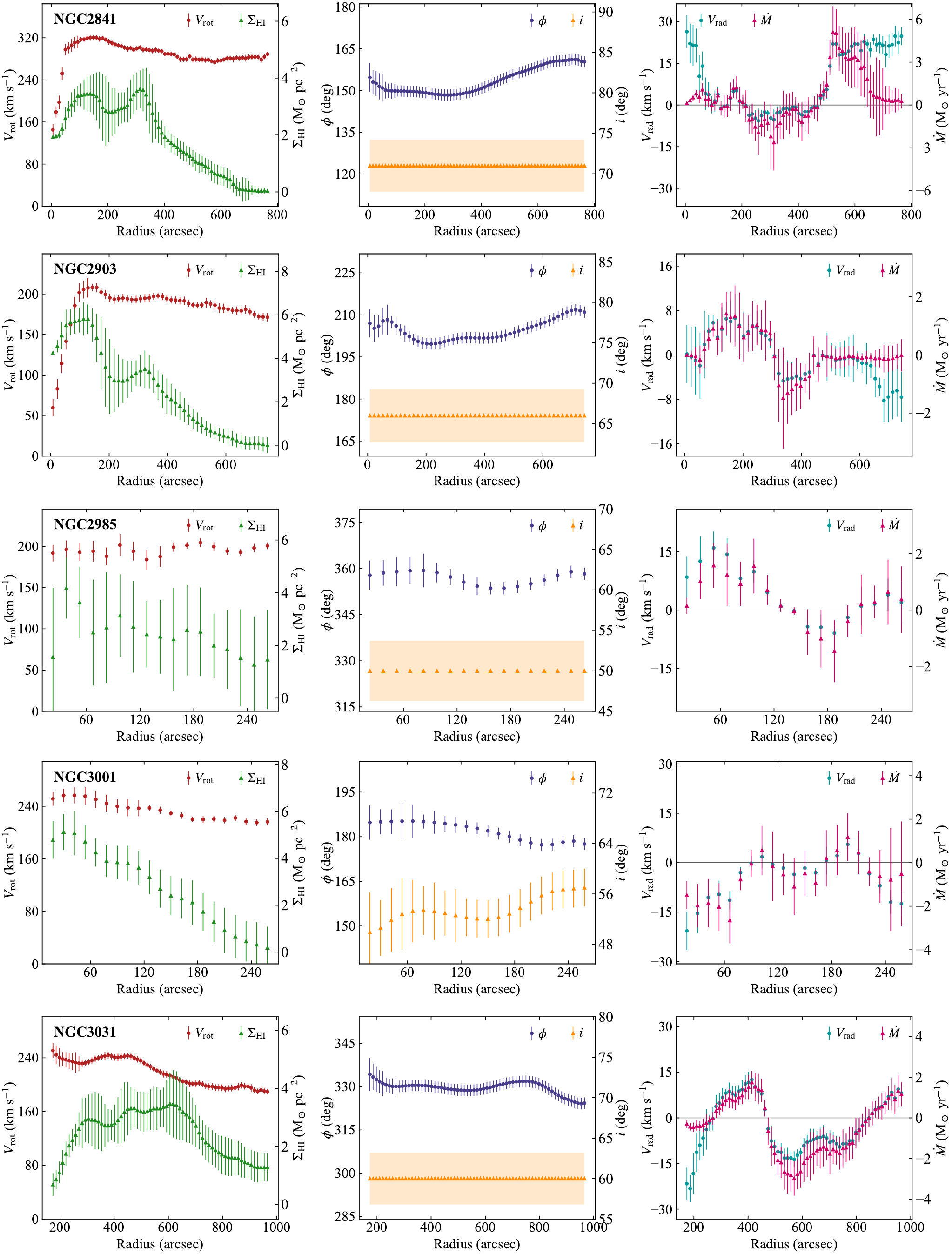}
    \caption{Continued}
\end{figure*}
\begin{figure*}
	\includegraphics[width=.99\textwidth]{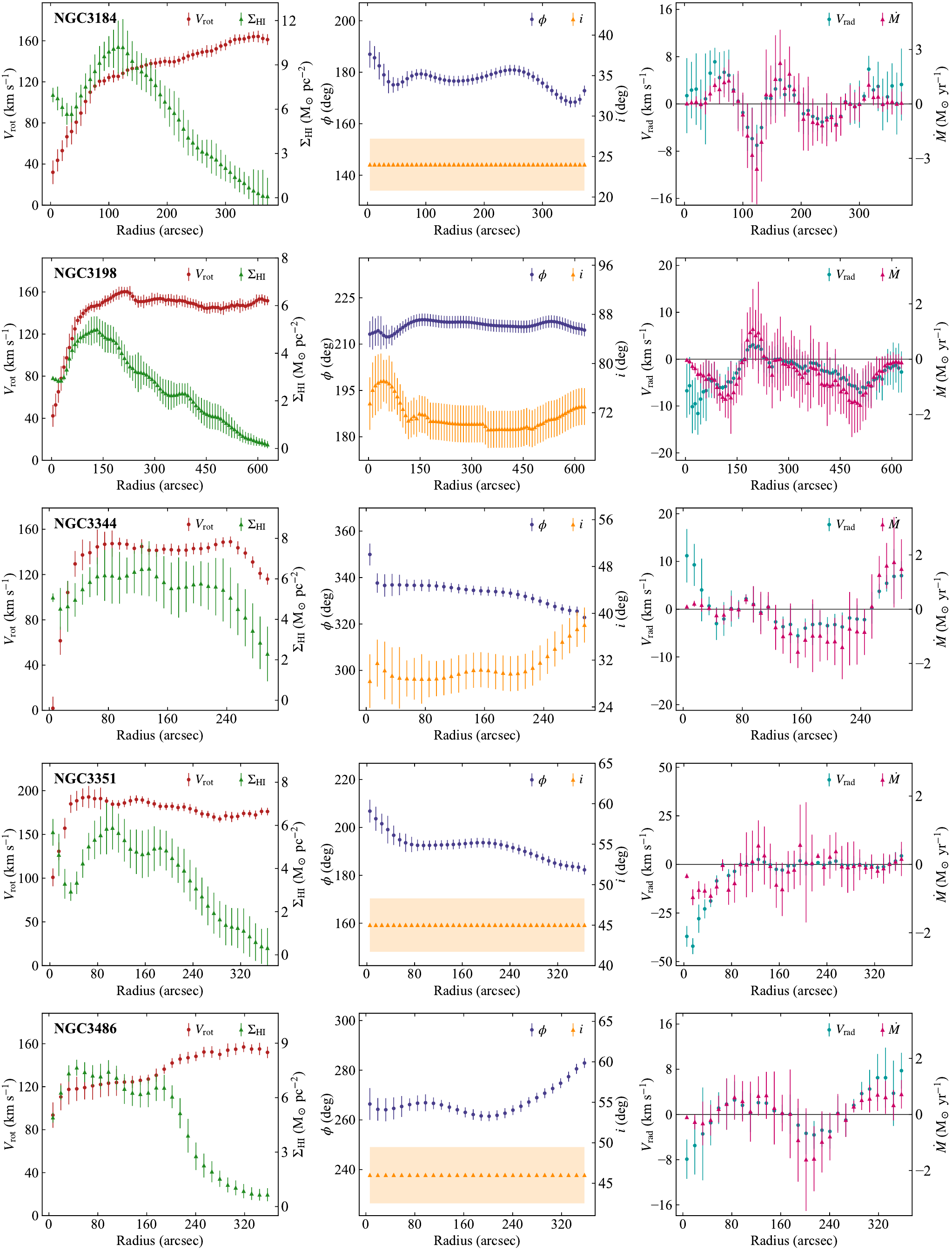}
    \caption{Continued}
\end{figure*}
\begin{figure*}
	\includegraphics[width=.99\textwidth]{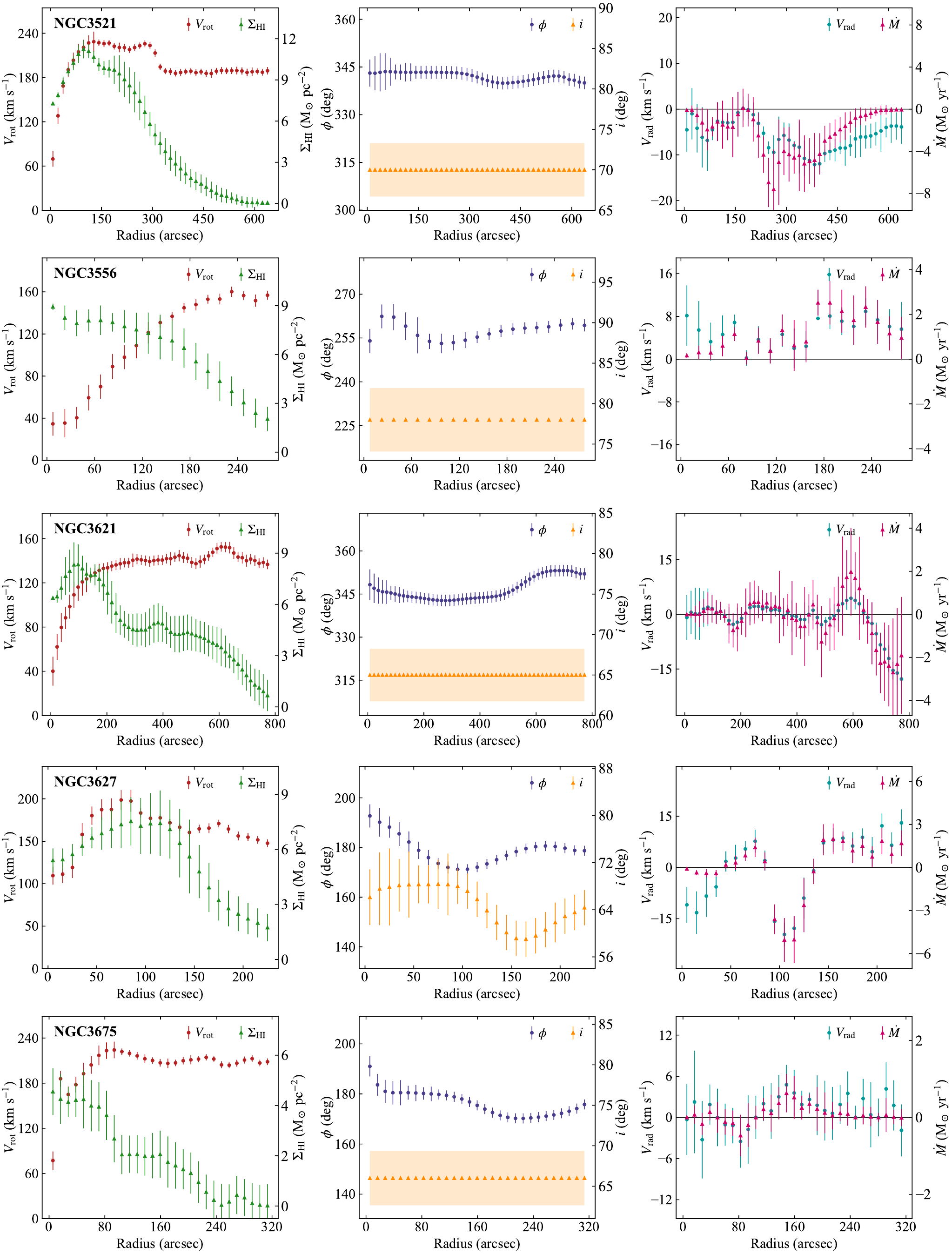}
    \caption{Continued}
\end{figure*}
\begin{figure*}
	\includegraphics[width=.99\textwidth]{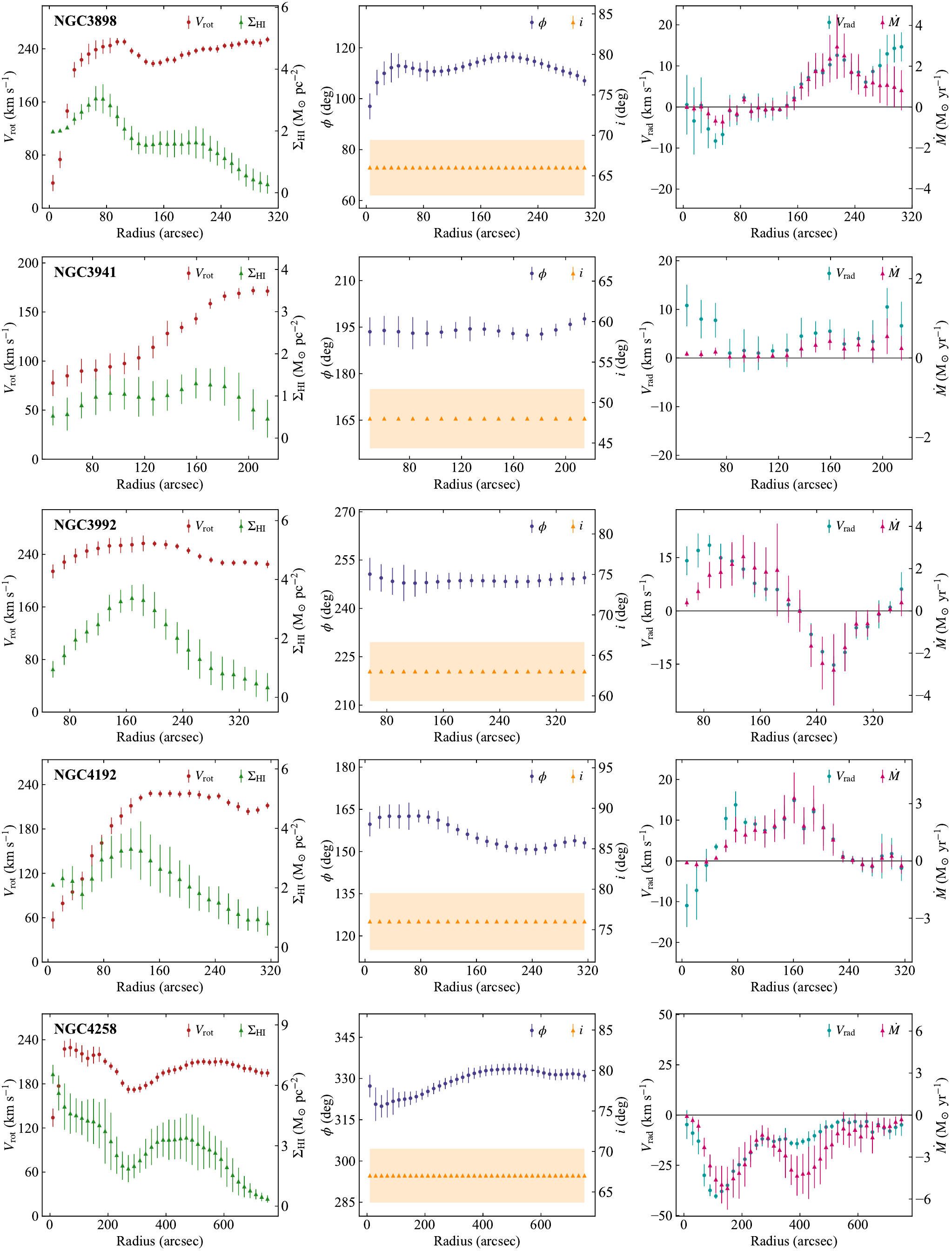}
    \caption{Continued}
\end{figure*}
\begin{figure*}
	\includegraphics[width=.99\textwidth]{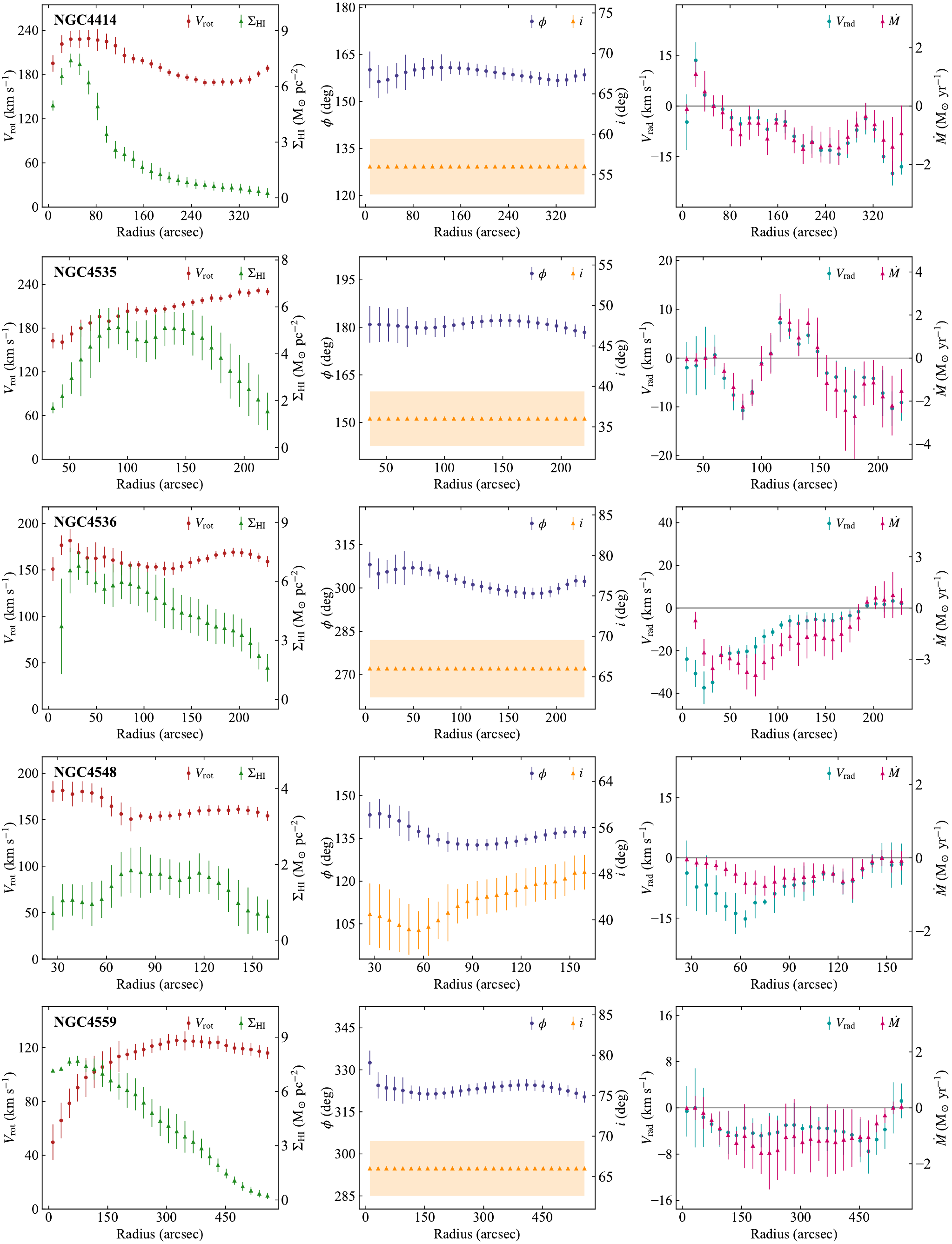}
    \caption{Continued}
\end{figure*}
\begin{figure*}
	\includegraphics[width=.99\textwidth]{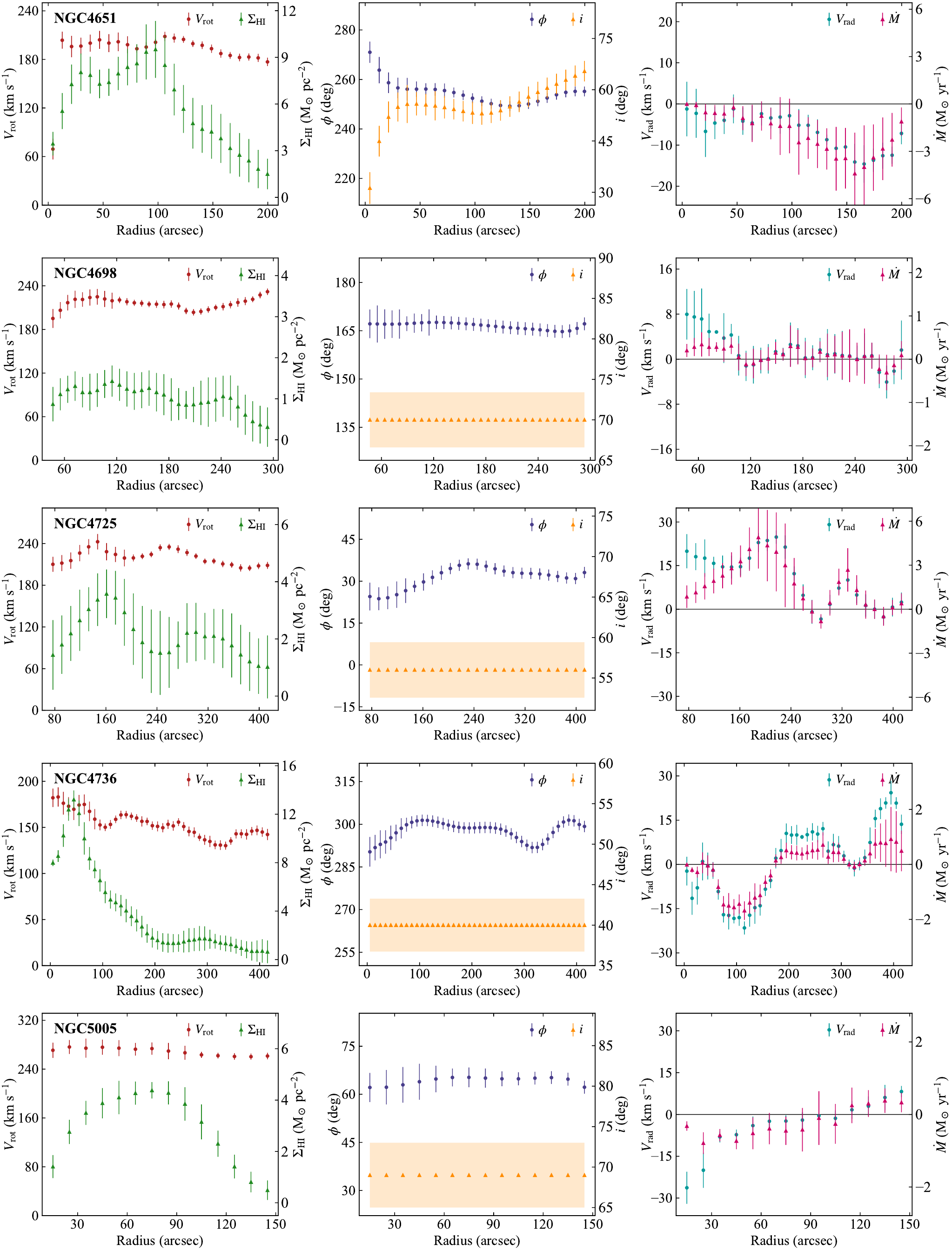}
    \caption{Continued}
\end{figure*}
\begin{figure*}
	\includegraphics[width=.99\textwidth]{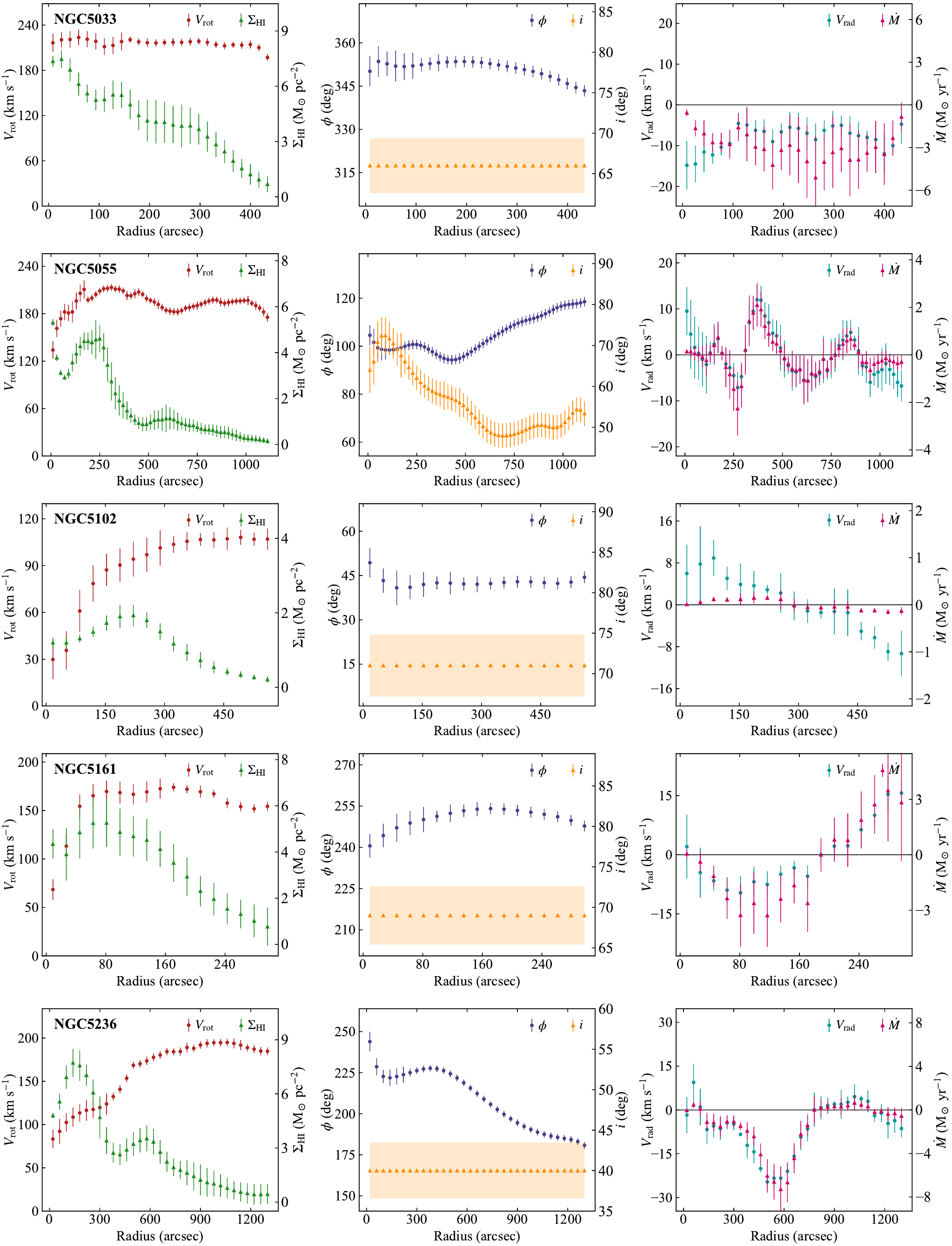}
    \caption{Continued}
\end{figure*}
\begin{figure*}
	\includegraphics[width=.99\textwidth]{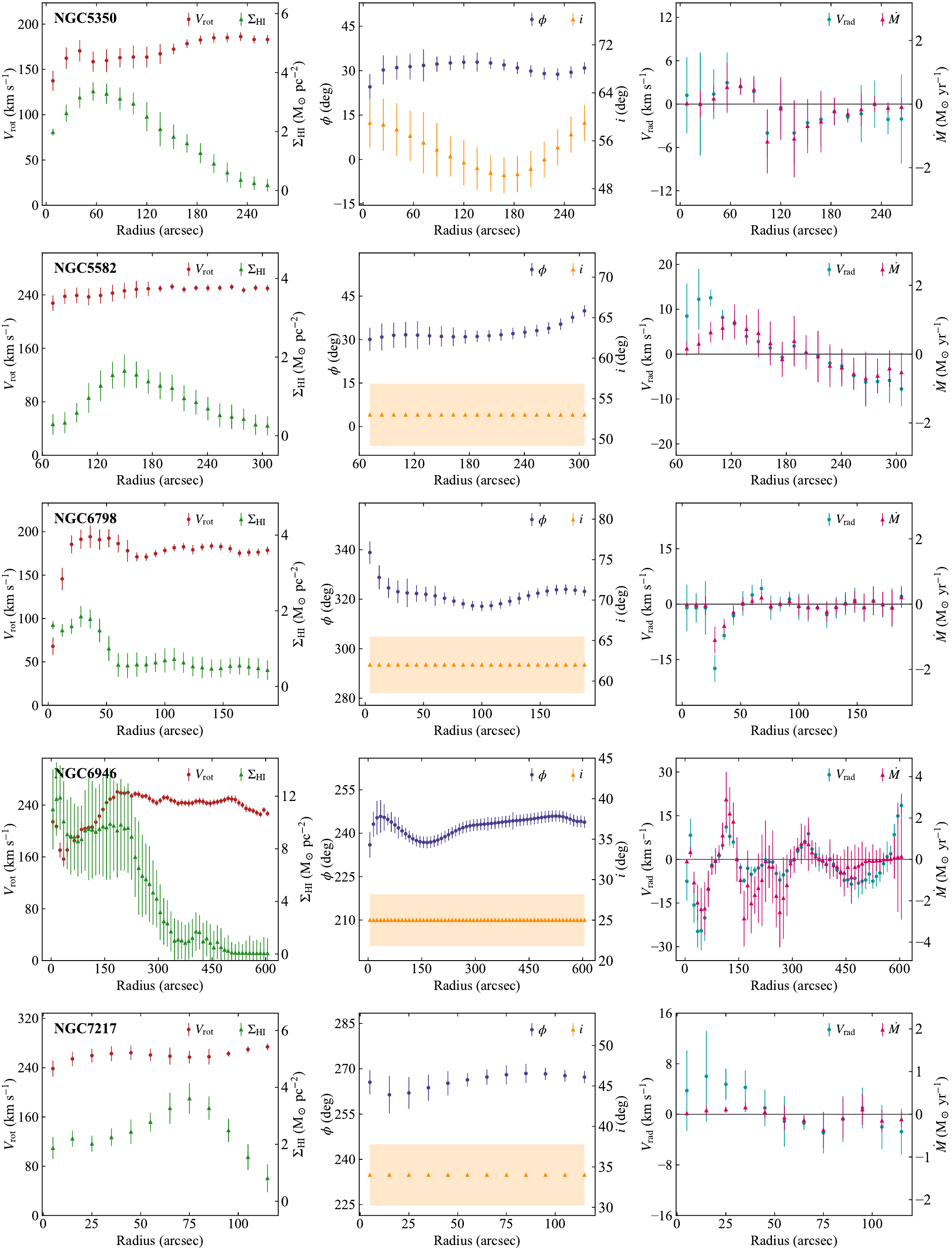}
    \caption{Continued}
\end{figure*}
\begin{figure*}
	\includegraphics[width=.99\textwidth]{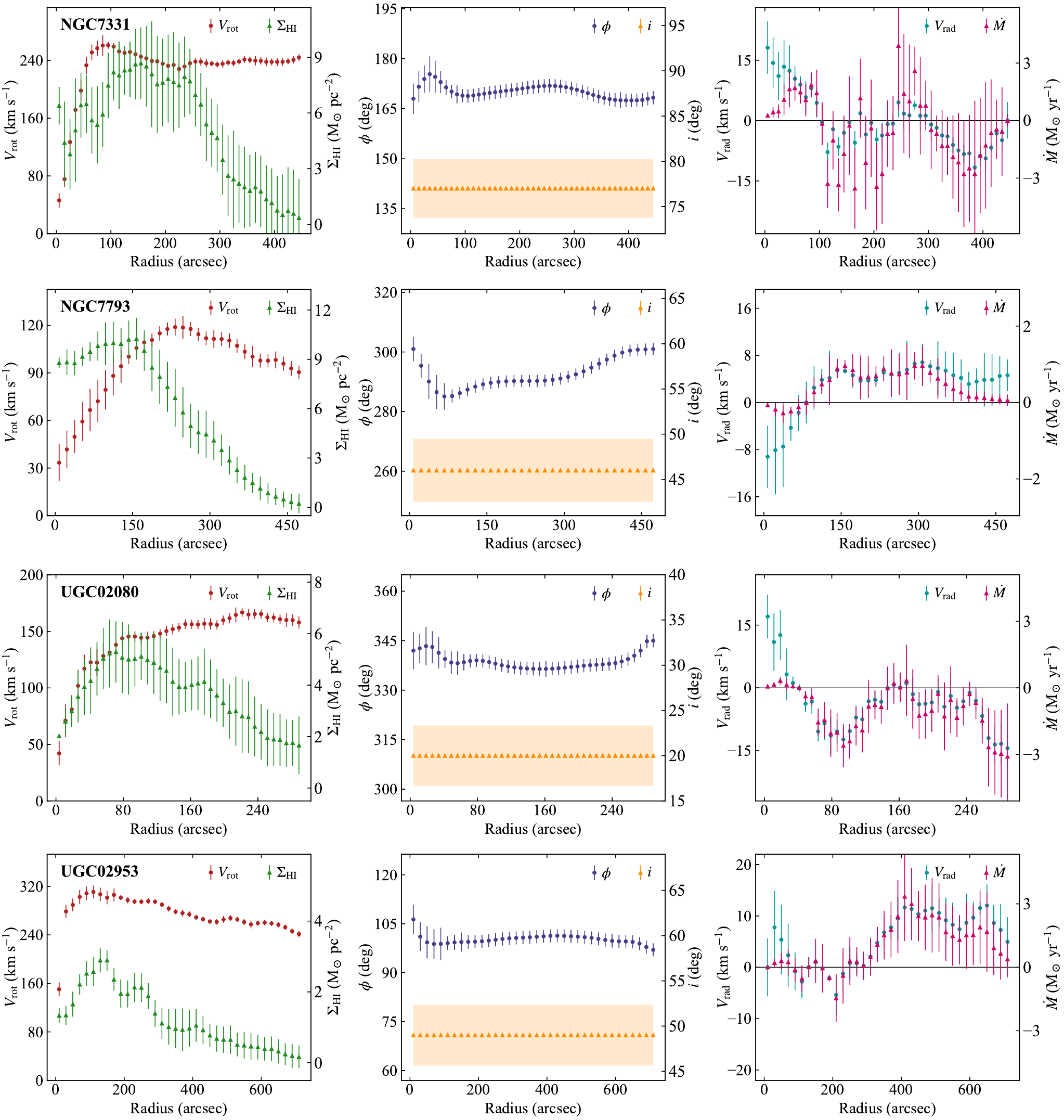}
    \caption{Continued}
\end{figure*}

\end{document}